\documentclass[aps,prb,reprint,superscriptaddress,floatfix,  longbibliography]{revtex4-2}

\usepackage{amsmath,amssymb,graphicx,bm}
\usepackage{physics}
\usepackage{hyperref}
\usepackage{xcolor}
\usepackage[percent]{overpic}

\begin{document}
\nocite{REVTEX42Control}

\title{Correlation-Driven Nonlinear Magnetoelectric Response in an Altermagnet: A Dynamical Mean-Field Study}

\author{Robert Peters}
\affiliation{%
Department of Physics, Kyoto University, Kyoto 606-8502, Japan}%
\email{peters@scphys.kyoto-u.ac.jp}

\author{Jun {\=O}ik\'{e}}
\affiliation{%
Department of Physics, Kyoto University, Kyoto 606-8502, Japan}

\date{\today}

\begin{abstract}
We investigate the optical nonlinear magnetoelectric effect (NMEE) in a
strongly correlated altermagnet using dynamical mean-field theory. Unlike
effective band descriptions with an imposed spin splitting, our approach
determines the altermagnetic order, electronic spectrum, and optical nonlinear response self-consistently. We find that the NMEE is finite in the
altermagnetic phase and vanishes in the paramagnetic phase. Its frequency
dependence reflects the spin-resolved spectral structure and provides an
estimate of the characteristic altermagnetic spin-splitting scale. Interaction
and temperature tuning produce qualitatively different behavior: at low
temperature, reducing the interaction strength toward the interaction-driven
magnetic phase boundary enhances the response, whereas increasing the temperature
suppresses it and drives it to zero above the critical temperature. These
results establish the optical NMEE as a probe of correlated altermagnetic order
and suggest that tuning parameters such as pressure, strain, or chemical
substitution toward an interaction-driven phase boundary may provide a
promising route to maximizing the response.
\end{abstract}

\maketitle

\section{Introduction}
Altermagnetism has recently emerged as a distinct class of collinear
magnetism, fundamentally different from both ferromagnetism and conventional
antiferromagnetism~\cite{PhysRevX.12.031042,PhysRevX.12.040501, Song2025AltermagnetsFunctionalMaterials,Jungwirth2026}. In a
ferromagnet, broken time-reversal symmetry ($\mathcal{T}$) produces a uniform
spin splitting and a finite macroscopic magnetization. In conventional
collinear antiferromagnets, the opposite-spin sublattices are related by a
translation or inversion symmetry, such that the bands remain spin degenerate.
In contrast, in an altermagnet the opposite-spin sublattices are connected by
a crystal rotation or a related point-group operation. As a result,
$\mathcal{T}$ is broken without generating a net magnetization, while the
electronic structure exhibits a momentum-dependent spin splitting with an
alternating sign across the Brillouin zone.

This momentum-resolved spin splitting often follows $d$-, $g$-, or $i$-wave
symmetry dictated by the crystal point group
~\cite{PhysRevX.12.031042,PhysRevX.12.040501,JPSJ.88.123702,
PhysRevB.102.144441}. Importantly, it does not require spin--orbit coupling,
but originates from nonrelativistic exchange interactions combined with the
symmetry relation between the opposite-spin sublattices. As a result,
altermagnets offer a platform for spin-dependent transport without stray
magnetic fields
~\cite{NatCommun.10.4305,PhysRevLett.126.127701,PhysRevX.12.011028}.
Experimentally, altermagnetic band splitting has been observed by
angle-resolved photoemission spectroscopy (ARPES) in candidate materials including
MnTe~\cite{PhysRevB.109.115102} and
CrSb~\cite{Yang2025CrSb}. 

Among the possible electrical signatures, the anomalous Hall effect has
attracted particular attention as a probe of altermagnetic order
~\cite{PhysRevB.102.075112,PhysRevB.106.195149,PhysRevB.110.094425,
PhysRevB.111.184408,8xv3-18nr}. However, it is not a generic consequence of
altermagnetic spin splitting: in many nonrelativistic altermagnets, magnetic
symmetries enforce a vanishing linear charge Hall conductivity. A finite
anomalous Hall effect therefore requires the relevant symmetry constraints to
be absent or lifted, for example by spin--orbit coupling or additional symmetry
breaking. This motivates complementary probes of the characteristic
momentum-dependent spin polarization.

The momentum-dependent spin polarization can also be understood in terms of
unconventional magnetic multipole order. The augmented-multipole framework
provides a systematic way to classify spin-split and reshaped band structures
in antiferromagnets without relying on spin--orbit coupling
~\cite{JPSJ.87.033709,JPSJ.88.123702,PhysRevB.102.144441}.
Within this framework, $d$-wave altermagnets have been identified as ferroic
phases whose leading time-reversal-odd order parameter is a magnetic octupole
rather than a dipole~\cite{PhysRevX.14.011019,PhysRevLett.132.176702}. Recent theoretical
developments have further formulated magnetic octupole moments in periodic
crystals from both quantum mechanical and thermodynamic perspectives,
clarifying their role as bulk order parameters in $d$-wave altermagnets
~\cite{SatoHayami2026NPJQuantumOctupole,frq1-9xx7,gxpm-2gkq,Sato2025OrbitalOctupole,Chen2026SpinMultipoles}.
This multipolar character imposes strong symmetry constraints on response
tensors and motivates probes such as piezomagnetism, which provides a
magnetoelastic signature of altermagnetic and octupolar order
~\cite{JPSJ.94.063704,PhysRevB.110.144421,PhysRevMaterials.8.L041402,Ma2021Piezomagnetism,JPSJ.94.083702,JPSJ.95.063701}.

Nonlinear electrical
responses offer a complementary class of probes, since they can be controlled
purely by electric fields and are strongly constrained by the same multipolar
symmetry. In particular, the nonlinear magnetoelectric effect (NMEE) was
recently proposed as a direct response signature of magnetic octupole order and
was shown to occur in both $d$-wave altermagnets and all-in/all-out magnetic
structures~\cite{PhysRevB.110.184407,PhysRevLett.129.086602,sxyb-5rcy,gly7-jzfl,Chatterjee2026NonlinearTellegen}. 
Third-order Hall responses have likewise been discussed as probes of
altermagnetic and octupolar order, both in altermagnetic systems and in
ferro-octupolar heavy-fermion compounds
~\cite{rv1n-vr4p,PhysRevLett.133.106701,PhysRevB.110.125127}.
Experimentally, giant room-temperature third-order electrical transport,
including a third-order Hall response, has recently been reported in
RuO$_2$ thin films, which were discussed as an altermagnet
candidate~\cite{Chen2026RuO2ThirdOrder}.
These developments establish
nonlinear response functions as powerful probes of altermagnetic and
octupolar order.

Much of the theoretical understanding of altermagnetism has been developed
through symmetry analysis and density functional theory (DFT), which have been used
to characterize spin-split band structures and identify candidate materials
~\cite{PhysRevX.12.031042,PhysRevX.12.040501,
Song2025AltermagnetsFunctionalMaterials,PhysRevB.110.144412}. These studies
have established the central role of magnetic space-group symmetries in
producing momentum-dependent spin splitting. More recently, interacting
lattice models have begun to address how altermagnetic order emerges and
evolves in correlated electron systems. Hartree--Fock and related weak-coupling
studies have demonstrated altermagnetic phases in Hubbard-type models
~\cite{PhysRevB.108.L100402,PhysRevLett.132.263402,
PhysRevB.111.L020401}, while quantum Monte Carlo and exact-diagonalization
approaches have explored fluctuation and strong-correlation effects beyond
static mean-field theory
~\cite{s31h-hk2v,Li2025DwavePairingAltermagnet,
Lu2025StonerBreakdownAltermagnetism}.
Complementary to these approaches, dynamical mean-field theory (DMFT) provides
access to real-frequency spectral functions and self-energies in a
self-consistent correlated magnetic state. DMFT calculations for
altermagnetic Hubbard models have shown that the dynamical self-energy can
qualitatively reshape the electronic spectrum beyond static mean-field
descriptions~\cite{7nvm-s225}. First-principles DFT+DMFT studies of candidate
materials such as CaCrO$_3$, NiS$_2$, and RuO$_2$ have likewise demonstrated
that electronic correlations, Hund's coupling, quasiparticle damping, and
incoherent spectral weight can play an essential role in determining their
electronic structure
~\cite{Ouyang2025CaCrO3,pgp6-zlh8,Park2026RuO2,
PhysRevB.99.184432}.

These many-body effects are expected to be particularly important for
nonlinear optical responses, which depend sensitively on the energies,
lifetimes, and spectral weights of the electronic excitations involved~\cite{PhysRevB.103.195133}. This
is especially relevant in altermagnets, where magnetic symmetries can suppress
conventional linear charge responses and thereby make nonlinear responses
important probes of the ordered state. Nevertheless, the influence of a
self-consistently determined correlated electronic structure on
optical nonlinear responses, and on the NMEE in particular,
remains largely unexplored.

In this work, we investigate how electronic correlations and temperature
control the nonlinear magnetoelectric response of an altermagnetic phase. To
this end, we study a minimal interacting lattice model within a self-consistent
DMFT framework. This approach allows us to
determine the magnetic order, the momentum-resolved spin polarization, and the
interacting spectral function on the same footing. We then evaluate the
optical NMEE using the resulting interacting Green's functions,
thereby directly connecting the nonlinear response to the correlated
spin-resolved spectral structure.
We show that the frequency dependence of the NMEE contains information about
the characteristic energy scale of the altermagnetic spin splitting, allowing
one to estimate a sector-averaged spin-splitting scale from the response peak
structure. We further demonstrate that tuning the interaction strength toward
the magnetic phase boundary enhances the NMEE, due to the reduced
gap and increased spin-polarized spectral weight near the Fermi level. In
contrast, increasing temperature suppresses the response through thermal
broadening and the weakening of altermagnetic order. These results establish
the NMEE as a sensitive probe of correlated altermagnetic order and its
underlying spin-resolved spectral properties.

The rest of this paper is organized as follows. In Sec.~\ref{sec:model}, we
introduce the interacting lattice model and describe the DMFT framework and the
response formulas used to evaluate the NMEE.
In Sec.~\ref{sec:results}, we present the main results. We first discuss the
spectral and response properties of a representative altermagnetic solution in
Sec.~\ref{subsec:reference_phase}. We then analyze the dependence on
interaction strength in Sec.~\ref{subsec:U_dependence}, followed by the
temperature dependence of the magnetic phase and nonlinear response in
Sec.~\ref{subsec:T_dependence}. Finally, Sec.~\ref{sec:conclusion} summarizes
our results.

\section{Model and Method}
\label{sec:model}
We consider an interacting lattice Hamiltonian
\begin{equation}
H = H_0 + H_{\mathrm{int}},
\end{equation}
where $H_0$ describes the noninteracting band structure and 
$H_{\mathrm{int}}$ denotes local Hubbard interactions.
The lattice contains two sublattices, labeled $A$ and $B$.
To describe the two-sublattice structure, we introduce Pauli matrices
$\bm{\tau}$ acting in sublattice space ($A/B$) and $\bm{\sigma}$
acting in spin space ($\uparrow/\downarrow$).
In the basis
\begin{equation}
\Psi_{\mathbf{k}}=
\bigl(
c_{\mathbf{k}A\uparrow},
c_{\mathbf{k}B\uparrow},
c_{\mathbf{k}A\downarrow},
c_{\mathbf{k}B\downarrow}
\bigr)^{\mathsf T},
\end{equation}
the single-particle Hamiltonian takes the form
\begin{equation}
H_0=\sum_{\mathbf{k}}
\Psi_{\mathbf{k}}^{\dagger}\,
\hat{H}_0(\mathbf{k})\,
\Psi_{\mathbf{k}},
\end{equation}
with
\begin{equation}
\hat{H}_0(\mathbf{k})
=
\Bigl[
\varepsilon_{0,\mathbf{k}}\,\tau_0
+t_{x,\mathbf{k}}\,\tau_x
+t_{z,\mathbf{k}}\,\tau_z
\Bigr]\otimes \sigma_0 ,
\label{eq:H0_pauli}
\end{equation}
where $\tau_0$ and $\sigma_0$ denote identity matrices in the
sublattice and spin spaces, respectively.
The dispersions are given by
\begin{align}
\varepsilon_{0,\mathbf{k}} &=E_0+
t_1 \left( \cos k_x + \cos k_y \right)  \notag\\
&\quad
+ t_2 \cos k_z
+ t_3 \cos k_x \cos k_y  \notag \\
&\quad
+ t_4 \left( \cos k_x + \cos k_y \right) \cos k_z \notag\\
&\quad
+ t_5 \cos k_x \cos k_y \cos k_z , \\
t_{x,\mathbf{k}} &=
t_8 \cos\frac{k_x}{2}\cos\frac{k_y}{2}\cos\frac{k_z}{2}, \\
t_{z,\mathbf{k}} &=
t_6 \sin k_x \sin k_y
+ t_7 \sin k_x \sin k_y \cos k_z .
\end{align}
Here, $E_0$ denotes the spin- and sublattice-independent onsite energy, while $t_i$ are hopping amplitudes. The parameters $t_1,\ldots,t_5$ describe sublattice-symmetric hopping contributions, $t_6$ and $t_7$ describe sublattice-asymmetric hopping contributions, and $t_8$ controls hopping between the two sublattices.
The interaction is an onsite Hubbard term on each sublattice,
\begin{equation}
H_{\mathrm{int}}
=
U \sum_{i}
\left(
n_{iA\uparrow} n_{iA\downarrow}
+
n_{iB\uparrow} n_{iB\downarrow}
\right),
\label{eq:Hint}
\end{equation}
with $n_{is\sigma}=c^{\dagger}_{is\sigma} c_{is\sigma}$ and $s\in\{A,B\}$.
Here, $U>0$ is the onsite Hubbard repulsion, representing the energy cost for simultaneous occupation of the same lattice site by an up-spin and a down-spin electron. The index $i$ labels the unit cells, and $A$ and $B$ denote the two sites within each unit cell.

The momentum-dependent hopping structure follows the tight-binding model
proposed in Ref.~\cite{PhysRevB.110.144412}, originally introduced in the
context of RuO$_2$. For the present study, we slightly adjust the hopping
parameters relative to the material-specific values of Ref.~\cite{PhysRevB.110.144412}
while preserving the symmetry structure responsible for altermagnetism,
in order to obtain a representative correlated altermagnetic insulator
at intermediate coupling suitable for the DMFT study. The key ingredient is the sublattice-asymmetric hopping term
$t_{z,\mathbf{k}}\tau_z$, whose leading momentum dependence
$\sin k_x\sin k_y$ changes sign under rotations in the $k_x$--$k_y$ plane.
When combined with staggered magnetic order between the two sublattices, this
term converts the otherwise conventional antiferromagnetic exchange field into
a momentum-dependent, sign-changing spin splitting characteristic of a
$d$-wave altermagnet. The inter-sublattice hopping $t_{x,\mathbf{k}}\tau_x$,
controlled mainly by $t_8$, hybridizes the two sublattices and sets the overall
bandwidth.

We express all energies in units of the noninteracting bandwidth $W$.
The dimensionless hopping parameters are $t_1/W=-0.0093$, $t_2/W=0.0130$, $t_3/W=0.0093$,
$t_4/W=-0.0278$, $t_5/W=-0.0074$, $t_6/W=-0.0185$, $t_7/W=0.0185$,
$t_8/W=0.5000$, $E_0/W=0.0463$.
The bare Hamiltonian is spin independent; the momentum-dependent spin
splitting discussed below arises from the self-consistently generated magnetic
self-energy.
Throughout this work, we consider the system at half-filling,
enforced by adjusting the chemical potential $\mu$ self-consistently
at each temperature and interaction strength.  We use units with $\hbar=1$, so that frequencies and energies are measured
in the same units. Accordingly, $\omega$, $\Omega$, $\varepsilon$, $T$,
$\eta$, and $W$ are all treated as energy scales in the numerical results.

We employ DMFT, which maps the lattice problem
onto a self-consistent quantum impurity model~\cite{RevModPhys.68.13}. The calculations are performed at temperature $T$. Within
DMFT, the electronic self-energy is local but retains its full frequency
dependence, allowing us to describe correlation-induced spectral
renormalization, quasiparticle damping, and finite-temperature effects beyond
static mean-field approaches.
The resulting impurity problem is solved using the numerical renormalization
group (NRG)~\cite{RevModPhys.47.773,RevModPhys.80.395}. This method provides
accurate Green's functions and self-energies directly on the real-frequency
axis and is particularly well suited for resolving the low-energy regime near
the Fermi level, which is essential for analyzing transport and low-frequency
response functions~\cite{PhysRevB.74.245114,PhysRevLett.99.076402}.

The interacting retarded and advanced lattice Green functions within DMFT are
given by
\begin{equation}
G^{R/A}(\mathbf{k},\omega)
=
\left[
(\omega+\mu\pm i\eta)\mathbb{I}
-\hat{H}_0(\mathbf{k})
-\Sigma^{R/A}(\omega)
\right]^{-1},
\label{eq:latticeGF}
\end{equation}
where the upper and lower signs correspond to the retarded and advanced Green
functions, respectively. Here, $\omega$ is the real-frequency variable, $\mu$
is the chemical potential adjusted self-consistently to maintain half filling,
and $\eta>0$ is an additional Green-function broadening used in the numerical
evaluation of the response functions.
If not stated differently, we here use $\eta/W=0.036$.
 The advanced quantities satisfy
$\Sigma^A(\omega)=[\Sigma^R(\omega)]^\dagger$ and
$G^A(\mathbf{k},\omega)=[G^R(\mathbf{k},\omega)]^\dagger$. 
In the ordered phase, $\Sigma(\omega)$ is a matrix in spin and sublattice
space. The altermagnetic solution is characterized by spin- and
sublattice-dependent local self-energies, with opposite local magnetizations on
the two sublattices. Thus, the momentum-dependent spin splitting arises from
the combination of the magnetic self-energy and the sublattice-dependent
hopping structure in Eq.~(\ref{eq:H0_pauli}).

From the interacting Green function in Eq.~(\ref{eq:latticeGF}), we define the
quantities used below to characterize the spectral and magnetic properties.
The matrix spectral function is
\begin{equation}
\hat A(\mathbf{k},\omega)
=
-\frac{1}{\pi}\,
\mathrm{Im}\,G^R(\mathbf{k},\omega),
\end{equation}
whose diagonal elements resolve the spin and sublattice components, while the
total momentum-resolved spectral function is
\begin{equation}
A(\mathbf{k},\omega)
=
\mathrm{Tr}\,\hat A(\mathbf{k},\omega).
\end{equation}
The local magnetization on sublattice $s\in\{A,B\}$ is defined as
\begin{equation}
m_s
=
\frac{1}{2}
\left(
\langle n_{s\uparrow}\rangle
-
\langle n_{s\downarrow}\rangle
\right),
\end{equation}
with
\begin{equation}
\langle n_{s\sigma}\rangle
=
\int d\omega\,f(\omega)\,
\frac{1}{N_k}\sum_{\mathbf{k}}
\left[\hat A(\mathbf{k},\omega)\right]_{s\sigma,s\sigma},
\end{equation}
where $f(\omega)$ is the Fermi distribution function. Finally, the
momentum-dependent spin polarization is characterized by
\begin{equation}
m(\mathbf{k})
=
\int d\omega\,f(\omega)
\left[
A_{\uparrow}(\mathbf{k},\omega)
-
A_{\downarrow}(\mathbf{k},\omega)
\right],
\end{equation}
where $A_\sigma(\mathbf{k},\omega)$ denotes the trace of
$\hat A(\mathbf{k},\omega)$ over the sublattice indices at fixed spin
$\sigma$.

Besides these static quantities, we use the optical conductivity and the NMEE
to characterize the altermagnetic state. 

We evaluate both responses using DMFT-dressed single-particle Green functions
within the bare-vertex approximation, neglecting two-particle vertex
corrections. For the momentum-independent self-energy employed here, the
resulting bare current vertices are consistent with the generalized Ward
identity~\cite{PhysRevB.103.195133,PhysRevB.110.165111}. Within this approximation, the current-current correlation
function is given by
\begin{align}
\Pi_{\alpha\beta}(\Omega)
=&
-\frac{1}{N_k}\sum_{\mathbf{k}}
\int \frac{d\varepsilon}{2\pi}\,
f(\varepsilon)\, \nonumber\\
&\times
\mathrm{Tr}\Bigl[
J_\alpha(\mathbf{k})
G^R(\mathbf{k},\varepsilon+\Omega)
J_\beta(\mathbf{k})
G^{R-A}(\mathbf{k},\varepsilon)
\nonumber\\
&\quad+
J_\alpha(\mathbf{k})
G^{R-A}(\mathbf{k},\varepsilon)
J_\beta(\mathbf{k})
G^A(\mathbf{k},\varepsilon-\Omega)
\nonumber\\
&\quad+
J_{\alpha\beta}(\mathbf{k})
G^{R-A}(\mathbf{k},\varepsilon)
\Bigr].
\end{align}
The superscripts $R$ and $A$ denote retarded and advanced Green functions, respectively, and $G^{R-A}\equiv G^R-G^A$.
The current and diamagnetic vertices are defined as
\begin{equation}
J_\alpha(\mathbf{k})
=
\frac{\partial \hat H_0(\mathbf{k})}{\partial k_\alpha},
\qquad
J_{\alpha\beta}(\mathbf{k})
=
\frac{\partial^2 \hat H_0(\mathbf{k})}
{\partial k_\alpha\,\partial k_\beta}.
\end{equation}
Here, $\Omega$ denotes the external optical frequency, while
$\alpha,\beta\in\{x,y,z\}$ label Cartesian spatial directions. The dissipative part of the optical conductivity is then obtained from
\begin{equation}
\sigma_{\alpha\beta}(\Omega)
=
\frac{1}{\Omega}
\mathrm{Im}\,
\Pi_{\alpha\beta}(\Omega).
\label{eq:sigma_real_frequency}
\end{equation}

The NMEE describes the spin polarization induced to second order by an
applied electric field,
\begin{equation}
\delta\langle s_\alpha(\omega_{\rm out})\rangle^{(2)}
=
\zeta^{(2)}_{\alpha;\beta\gamma}
(\omega_{\rm out};\omega_1,\omega_2)
E_\beta(\omega_1)E_\gamma(\omega_2),
\end{equation}
where $\omega_{\rm out}=\omega_1+\omega_2$. 

The NMEE is obtained from the second-order Kubo formula evaluated using the
interacting Green functions in Eq.~(\ref{eq:latticeGF})
~\cite{PhysRevB.110.165111,PhysRevB.103.195133,PhysRevB.110.184407,PhysRevB.94.144432,gly7-jzfl}.
The corresponding response kernel is given by
\begin{widetext}
\begin{align}
\kappa^{(2)}_{\alpha;\beta\gamma}
(\omega_{\rm out};\omega_1,\omega_2)
&=
\frac{1}{N_k}\sum_{\mathbf{k}}
\int_{-\infty}^{\infty}
\frac{d\varepsilon}{2\pi i}\,
f(\varepsilon)\,
\mathrm{Tr}\Bigg[
\nonumber\\
&\quad
\frac{1}{2}
s_\alpha
G^R(\mathbf{k},\varepsilon+\omega_{\rm out})
J_{\beta\gamma}(\mathbf{k})
G^{R-A}(\mathbf{k},\varepsilon)
\nonumber\\
&\quad
+
\frac{1}{2}
s_\alpha
G^{R-A}(\mathbf{k},\varepsilon)
J_{\beta\gamma}(\mathbf{k})
G^A(\mathbf{k},\varepsilon-\omega_{\rm out})
\nonumber\\
&\quad
+
s_\alpha
G^R(\mathbf{k},\varepsilon+\omega_{\rm out})
J_\beta(\mathbf{k})
G^R(\mathbf{k},\varepsilon+\omega_2)
J_\gamma(\mathbf{k})
G^{R-A}(\mathbf{k},\varepsilon)
\nonumber\\
&\quad
+
s_\alpha
G^R(\mathbf{k},\varepsilon+\omega_1)
J_\beta(\mathbf{k})
G^{R-A}(\mathbf{k},\varepsilon)
J_\gamma(\mathbf{k})
G^A(\mathbf{k},\varepsilon-\omega_2)
\nonumber\\
&\quad
+
s_\alpha
G^{R-A}(\mathbf{k},\varepsilon)
J_\beta(\mathbf{k})
G^A(\mathbf{k},\varepsilon-\omega_1)
J_\gamma(\mathbf{k})
G^A(\mathbf{k},\varepsilon-\omega_{\rm out})
\Bigg]
\nonumber\\
&\quad
+
\big[(\beta,\omega_1)
\leftrightarrow
(\gamma,\omega_2)\big].
\label{eq:nmee_tensor}
\end{align}
\end{widetext}
Here, $s_\alpha$ denotes the spin operator, and
$[(\beta,\omega_1)\leftrightarrow(\gamma,\omega_2)]$ indicates symmetrization
with respect to the two external fields. We focus on the rectification-type
frequency configuration
\[
  \omega_1=\Omega,\qquad \omega_2=-\Omega,\qquad \omega_{\rm out}=0,
\]
and on the tensor component
\begin{equation}
  \kappa_{z,xy}(\Omega)
  \equiv
  \kappa^{(2)}_{z;xy}(0;\Omega,-\Omega).
\end{equation}
For the symmetry of the present model, the real part of this tensor component remains finite, while the imaginary part vanishes. 
The finite real part describes a rectified spin polarization induced by linearly polarized light, whose magnitude and sign depend on the polarization angle. The physical response tensor is then obtained as
\begin{equation}
\zeta^{(2)}_{\alpha;\beta\gamma}
(\omega_{\rm out};\omega_1,\omega_2)=
\frac{1}
{(\omega_1+i0^+)(\omega_2+i0^+)}
\kappa^{(2)}_{\alpha;\beta\gamma}.
\end{equation}
Here, $i0^+$ denotes a positive infinitesimal that specifies the causal response and corresponds to adiabatic switching on of the external field. 
Thus, for the frequency configuration used below, the measurable response is proportional to $\mathrm{Re}\,\kappa_{z,xy}(\Omega)/\Omega^2$, up to the overall sign and the infinitesimal convergence factors.

\section{Results}
\label{sec:results}
\begin{figure*}[t]
\centering 
\begin{overpic}[width=0.32\linewidth] {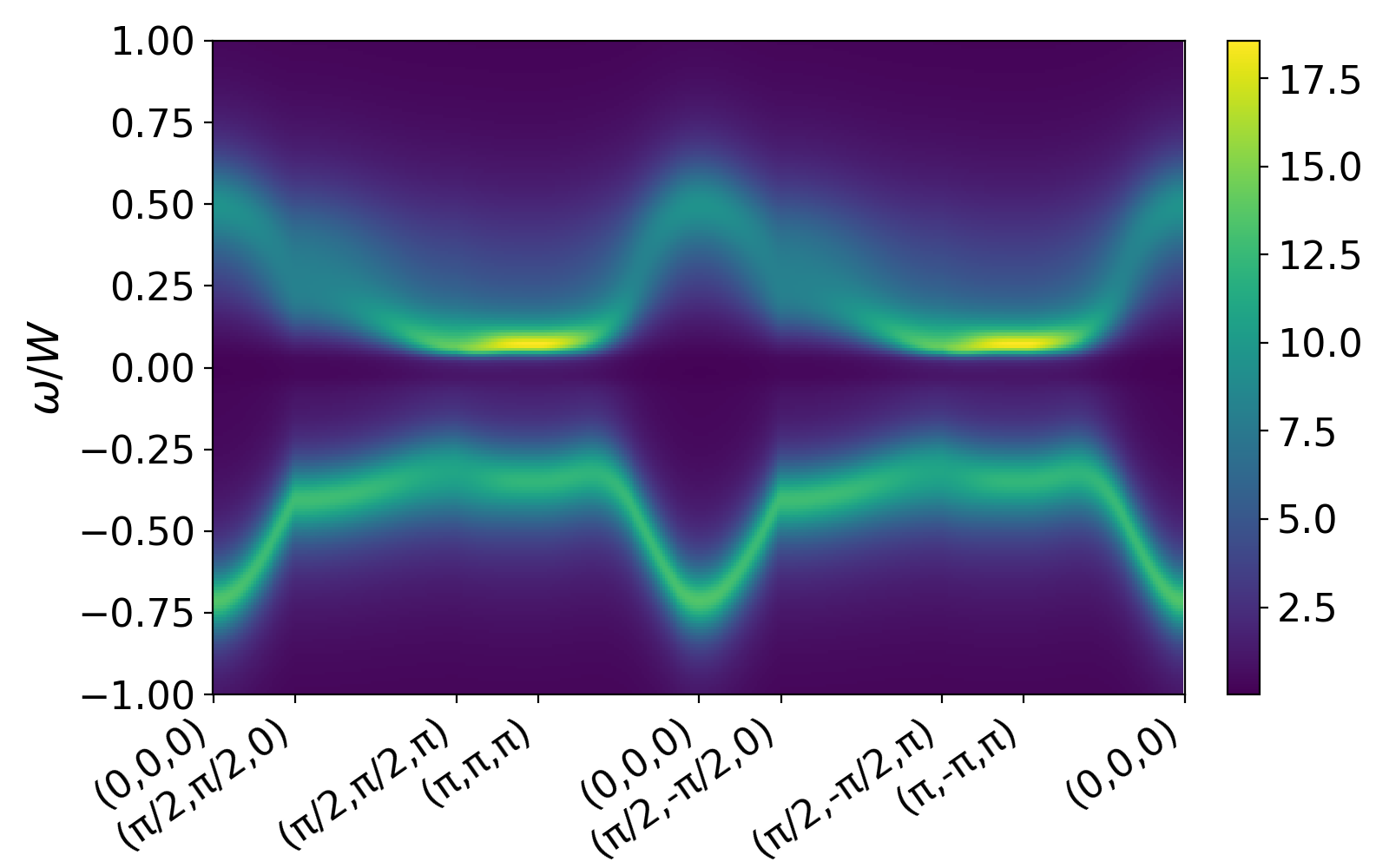} 
\put(0,60){(a)} 
\end{overpic} 
\hfill 
\begin{overpic}[width=0.32\linewidth] {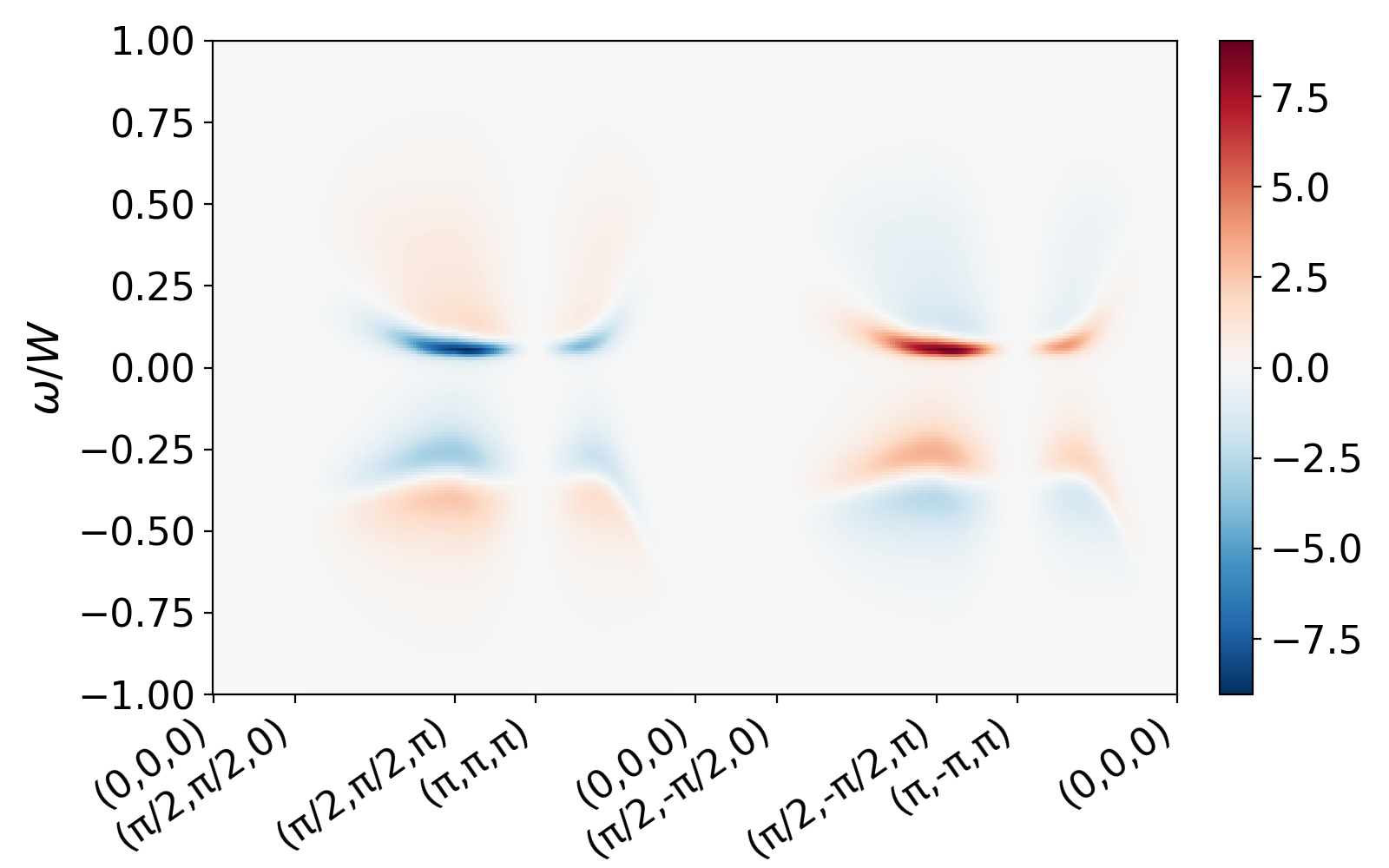}
 \put(0,60){(b)}
  \end{overpic} \hfill \begin{overpic}[width=0.32\linewidth] {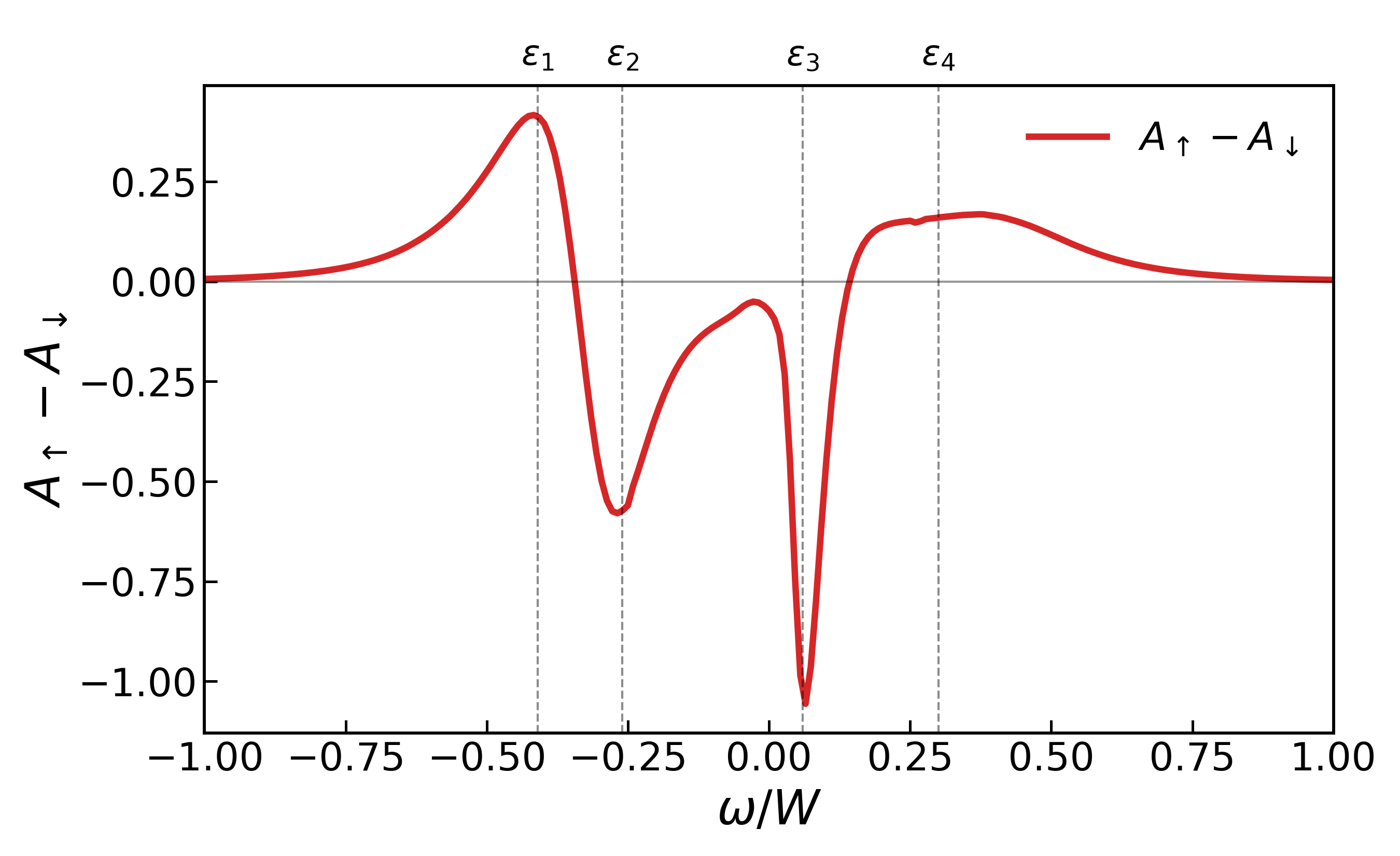} 
  \put(0,60){(c)} 
  \end{overpic}
\caption{
Spectral properties in the altermagnetic phase at $U/W=0.75$ and
$T/W=0.001$.
(a): Momentum-resolved spectral function $A_\uparrow + A_\downarrow$.
(b): Momentum-resolved spin difference $A_\uparrow - A_\downarrow$,
revealing the characteristic momentum-dependent spin polarization.
The sign change under a $90^\circ$-rotation in the $k_x$--$k_y$ plane reflects the
$d$-wave-like altermagnetic pattern.
(c) Sector-integrated spin difference obtained by integrating over
$k_x,k_y\in[0,\pi]$ and $k_z\in[-\pi,\pi]$. This restricted momentum
integration avoids the full Brillouin-zone cancellation associated with the
$d$-wave-like spin texture and allows us to estimate a characteristic
sector-averaged spin-splitting scale. The four dominant peak positions are
marked at $\epsilon_1/W\simeq-0.41$, $\epsilon_2/W\simeq-0.26$,
$\epsilon_3/W\simeq0.06$, and $\epsilon_4/W\simeq0.30$.
\label{fig_specU04}
}
\end{figure*}
\subsection{Altermagnetic phase: spectral and response properties}
\label{subsec:reference_phase}

\begin{figure*}[t]
\centering 
\begin{overpic}[width=0.32\linewidth]{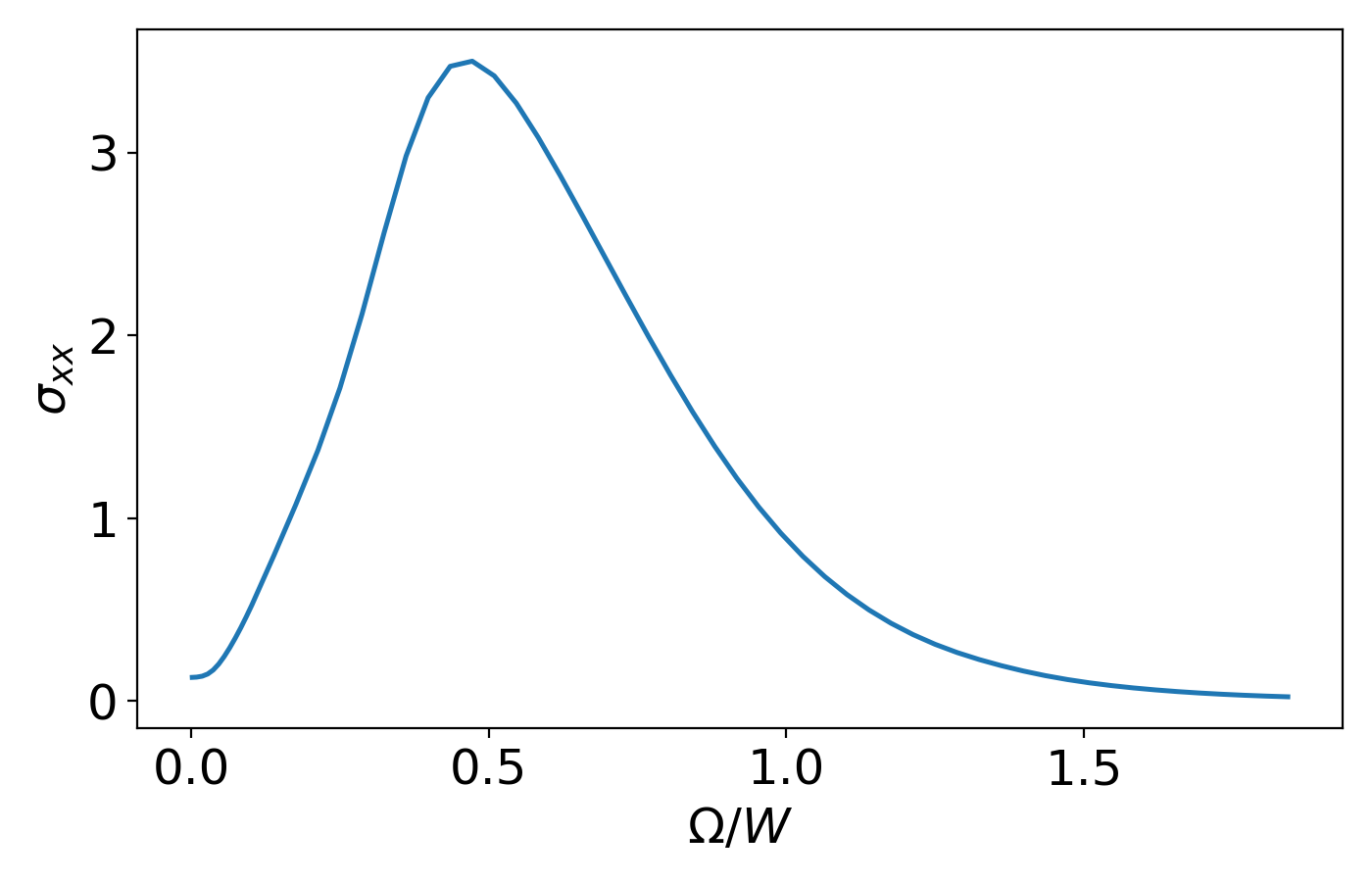}
\put(0,63){(a)} 
\end{overpic} 
\hfill 
\begin{overpic}[width=0.32\linewidth]{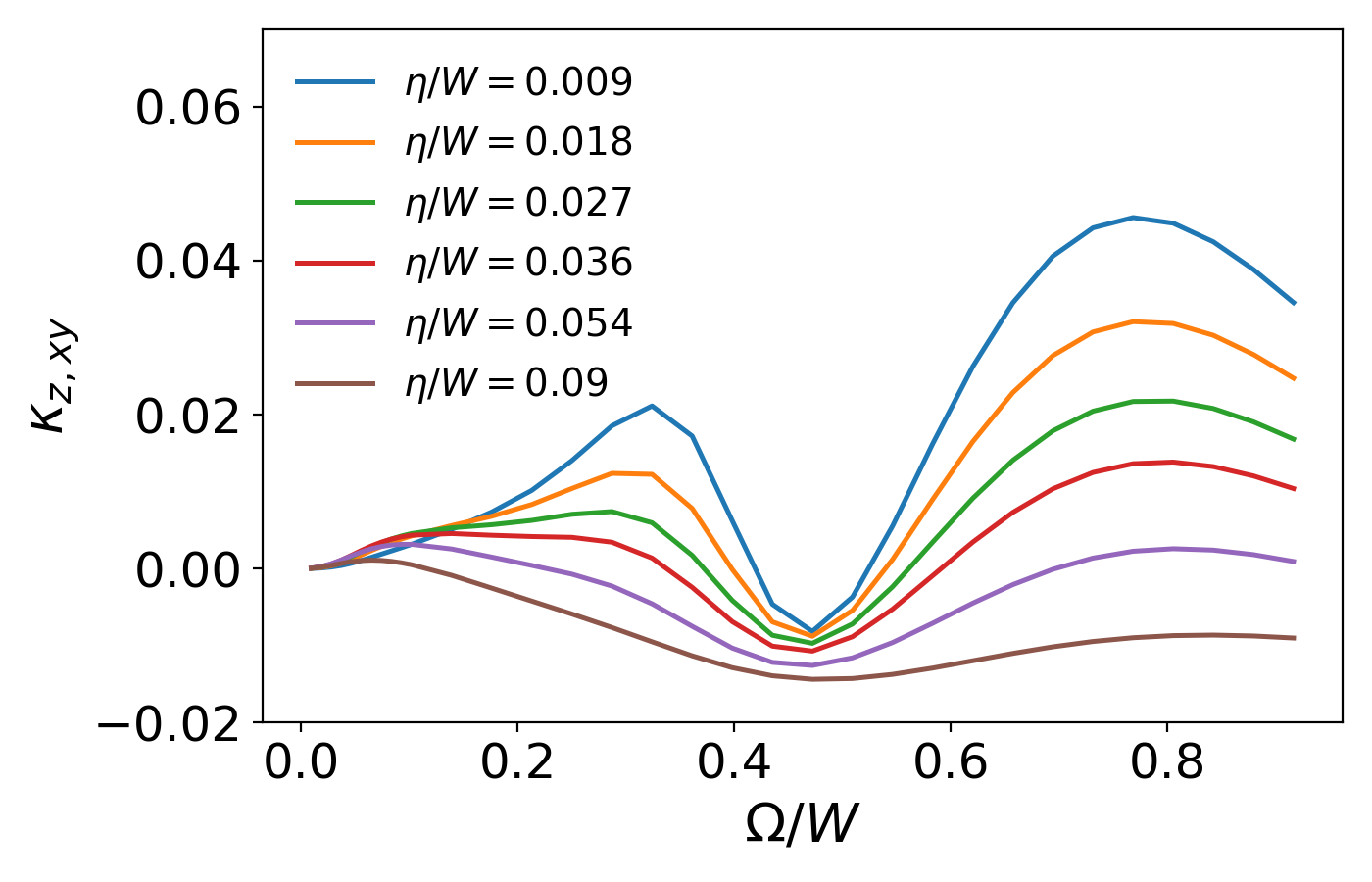}
\put(0,63){(b)} 
\end{overpic} 
\hfill 
\begin{overpic}[width=0.32\linewidth]{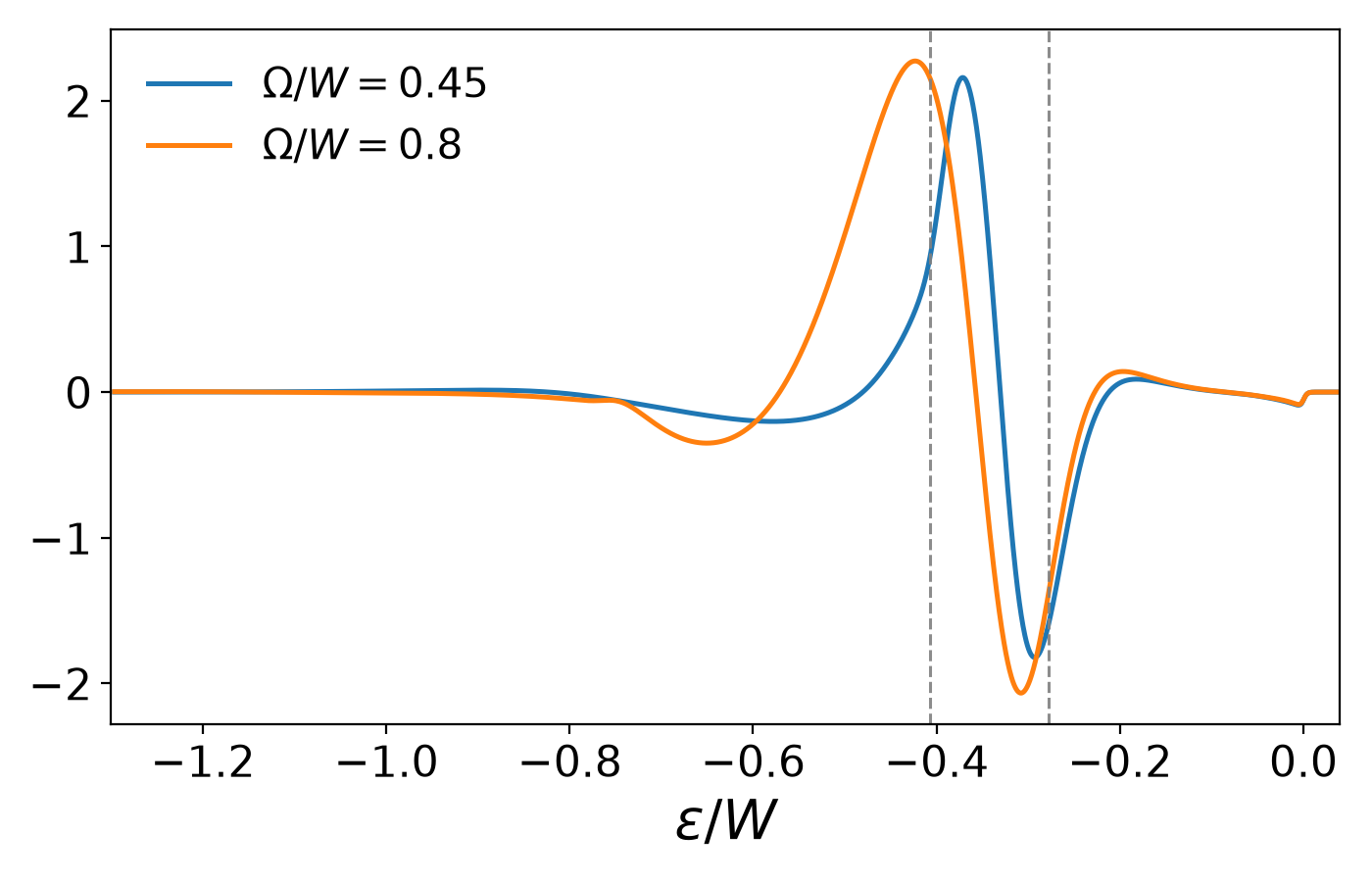}
\put(0,63){(c)} 
\end{overpic} 
\caption{Linear and nonlinear optical responses for $U/W=0.75$ in the altermagnetic
phase. (a) Longitudinal optical conductivity $\sigma_{xx}(\Omega)$, shown as a
baseline charge response. (b) Real part of the optical NMEE,
$\mathrm{Re}\,\kappa_{z,xy}(\Omega)$, for different Green-function
broadenings $\eta/W$. (c) Energy-resolved integrand of
$\mathrm{Re}\,\kappa_{z,xy}(\Omega)$} as a function of the internal energy
$\varepsilon$, evaluated at the two higher-frequency extrema,
$\Omega/W=0.45$ and $0.8$. The vertical dashed lines mark the occupied
spin-polarized spectral features at $\varepsilon_1/W\simeq-0.41$ and
$\varepsilon_2/W\simeq-0.26$. The full response is obtained by integrating the
plotted quantity over $\varepsilon$.
\label{fig_responseU04}
\end{figure*}

We first establish the spectral properties of the altermagnetic phase, shown in
Fig.~\ref{fig_specU04}. The momentum-resolved total spectral function
$A_\uparrow + A_\downarrow$ [Fig.~\ref{fig_specU04}(a)] exhibits a clear gap structure around the Fermi
energy, $\omega=0$, while the momentum-resolved spin difference
$A_\uparrow - A_\downarrow$ [Fig.~\ref{fig_specU04}(b)] reveals a momentum-dependent spin polarization.
Upon a $90^\circ$-rotation in the $k_x$--$k_y$ plane, the sign of this polarization changes,
consistent with a $d$-wave-like altermagnetic pattern. This momentum-dependent spin splitting is the characteristic microscopic feature underlying the
nonlinear magnetoelectric response discussed below.

In addition to the momentum-resolved spectra, Fig.~\ref{fig_specU04}(c) shows
the partially momentum-integrated spin contrast. Since the altermagnetic spin
polarization changes sign across the Brillouin zone, a full momentum
integration would cancel the spin-resolved signal. We therefore integrate over
the momentum sector $k_x,k_y\in[0,\pi]$ and $k_z\in[-\pi,\pi]$, where the spin
polarization has a fixed sign. The sector-integrated spectrum reveals four
prominent spin-polarized spectral features, which we denote by $\varepsilon_i$
in order of increasing energy:
$\varepsilon_1/W\approx-0.41$, $\varepsilon_2/W\approx-0.26$,
$\varepsilon_3/W\approx0.06$, and $\varepsilon_4/W\approx0.30$.
These characteristic energies will be used below to interpret the peak
structure of the NMEE.

The spin splittings of the occupied and unoccupied spectral manifolds are not
identical. The two occupied features yield
$\Delta_{\rm sp}^{\rm occ}\simeq\varepsilon_2-\varepsilon_1\approx0.15W$,
whereas the two unoccupied features yield
$\Delta_{\rm sp}^{\rm unocc}\simeq\varepsilon_4-\varepsilon_3\approx0.24W$.
Such an asymmetry is natural in the correlated altermagnetic state. In
contrast to an effective band description with an imposed, energy-independent
spin splitting, the DMFT self-energy is frequency dependent. Its real part
shifts and renormalizes spectral features differently at different energies,
while its imaginary part produces an energy-dependent broadening. The
occupied and unoccupied spin-polarized manifolds therefore need not exhibit
the same effective splitting or linewidth. We define a characteristic
sector-averaged spin-splitting scale as
$\overline{\Delta}_{\rm sp}
=(\Delta_{\rm sp}^{\rm occ}+\Delta_{\rm sp}^{\rm unocc})/2
\approx0.195W$. Because the highest-energy feature at $\varepsilon_4$ is
relatively broad, the averaged value should
be regarded as approximate characteristic energy scales of the interacting
spectral structure rather than sharply defined band splittings.

Having established the spectral properties of the altermagnetic phase, we now
analyze the corresponding optical response functions shown in
Fig.~\ref{fig_responseU04}. We first consider the longitudinal optical
conductivity $\sigma_{xx}(\Omega)$ as a baseline charge response. This allows
us to test whether a conventional linear optical probe already carries a clear
signature of the altermagnetic spin texture, before turning to the
spin-sensitive nonlinear response. The optical conductivity
$\sigma_{xx}(\Omega)$ [Fig.~\ref{fig_responseU04}(a)] exhibits a broad peak
around $\Omega/W\approx0.46$ and is strongly suppressed at low frequencies,
consistent with the gapped electronic state visible in the spectral
function. The small residual low-frequency conductivity originates from the
finite Green-function broadening $\eta/W$. Apart from confirming the
insulating character of the solution, however, the longitudinal optical
conductivity does not provide a clear signature of the sign-changing
altermagnetic spin polarization. This motivates the analysis of the optical
NMEE as a spin-sensitive nonlinear response.

We next consider the optical NMEE. Although the measurable response is
proportional to $\mathrm{Re}\,\kappa_{z,xy}(\Omega)/\Omega^2$, we plot
$\mathrm{Re}\,\kappa_{z,xy}(\Omega)$ itself in Fig.~\ref{fig_responseU04}(b) for several
values of the Green-function broadening $\eta/W$. This representation more
clearly reveals the underlying transition structure, since
$\kappa_{z,xy}(\Omega)$ can be related directly to transitions between
spin-polarized spectral regions. By contrast, the additional factor
$1/\Omega^2$ strongly enhances the low-frequency region and obscures the
higher-frequency structures relevant for identifying characteristic excitation
energies.
The low-frequency positive feature depends strongly on $\eta/W$ and occurs in
a regime where the external frequency is comparable to the numerical
broadening. We therefore do not assign a universal optical excitation scale to
this feature. Instead, we focus on the two higher-frequency extrema: a negative
local minimum near $\Omega/W\simeq0.45$ and a positive maximum near
$\Omega/W\simeq0.8$. Their positions depend only weakly on the broadening and
can be related to transitions between spin-polarized spectral regions.

To clarify the spectral origin of the higher-frequency structures,
Fig.~\ref{fig_responseU04}(c) shows the energy-resolved integrand at the
corresponding external frequencies. The full response is obtained by
integrating this quantity over the internal energy $\varepsilon$, so the
observed sign and magnitude result from cancellations between positive and
negative contributions. For the lower-frequency local minimum at
$\Omega/W\simeq0.45$, the integrand contains sizable contributions of both
signs. The negative-signed weight near the occupied spin-contrast feature
$\varepsilon_2/W\simeq-0.26$ slightly outweighs the positive contribution near
$\varepsilon_1/W\simeq-0.41$, producing the small negative extremum observed in
Fig.~\ref{fig_responseU04}(b). For the higher-frequency maximum at
$\Omega/W\simeq0.8$, the positive contribution near
$\varepsilon_1/W\simeq-0.41$ is enhanced and gives rise to the much
larger positive NMEE peak, although a negative contribution near
$\varepsilon_2$ remains present.

This energy-resolved structure can be related to the underlying optical
transitions as follows. Since the Hamiltonian and current operators are spin
diagonal, the relevant optical transitions conserve spin. A contribution at
internal energy $\varepsilon$ is enhanced when appreciable occupied spectral
weight at $\varepsilon$ is connected by the external frequency $\Omega$ to
unoccupied spectral weight near $\varepsilon+\Omega$ in the same spin channel.
In this sense, the lower-frequency local minimum is associated with the inner
transition manifold involving the regions near $\varepsilon_2$ and
$\varepsilon_3$, whereas the higher-frequency maximum is associated with the
outer transition manifold involving the regions near $\varepsilon_1$ and the
broad upper unoccupied feature around $\varepsilon_4/W\simeq0.30$. Because both extrema contain sizable positive and negative contributions from
multiple occupied energy regions, they should not be interpreted as unique
transitions between individual spectral peaks. They are more appropriately
understood as weighted transition manifolds whose net signs and magnitudes are
set by the balance of competing contributions in the NMEE kernel.

Within this interpretation, the lower-frequency negative extremum is predominantly
associated with the inner transition manifold connecting the regions near
$\varepsilon_2$ and $\varepsilon_3$, whereas the higher-frequency extremum is
predominantly associated with the outer manifold connecting the regions near
$\varepsilon_1$ and $\varepsilon_4$. Their separation,
$\Delta\Omega/W\simeq0.8-0.45\simeq0.35$, therefore approximately measures the
difference between the outer and inner transition scales. For the corresponding
spectral energies,
\begin{align}
(\varepsilon_4-\varepsilon_1)-(\varepsilon_3-\varepsilon_2)
&=
(\varepsilon_2-\varepsilon_1)+(\varepsilon_4-\varepsilon_3)
\nonumber\\
&=
\Delta_{\mathrm{sp}}^{\mathrm{occ}}
+
\Delta_{\mathrm{sp}}^{\mathrm{unocc}}
=
2\overline{\Delta}_{\mathrm{sp}}.
\end{align}
This motivates using half the NMEE peak separation as an optical estimate of
the average spin-splitting scale,
$\overline{\Delta}_{\mathrm{sp}}^{\mathrm{NMEE}}/W\simeq\Delta\Omega/(2W)\simeq0.175$.
This value is somewhat smaller than, but comparable to, the sector-averaged
estimate $\overline{\Delta}_{\mathrm{sp}}/W\simeq0.195$ obtained from the
occupied and unoccupied peak separations in
Fig.~\ref{fig_specU04}(c). Exact agreement is not expected because the NMEE
extrema represent broad momentum- and matrix-element-weighted transition
manifolds, whereas the spectral estimate is extracted from a restricted
momentum sector. The broad upper unoccupied feature introduces additional
uncertainty into the spectral estimate. Nevertheless, the NMEE frequency
structure provides a semiquantitative, indirect optical estimate of the
characteristic altermagnetic spin-splitting scale.


\subsection{Interaction strength dependence of the magnetic phase and nonlinear response}
\label{subsec:U_dependence}
\begin{figure}[t]
\includegraphics[width=\linewidth]{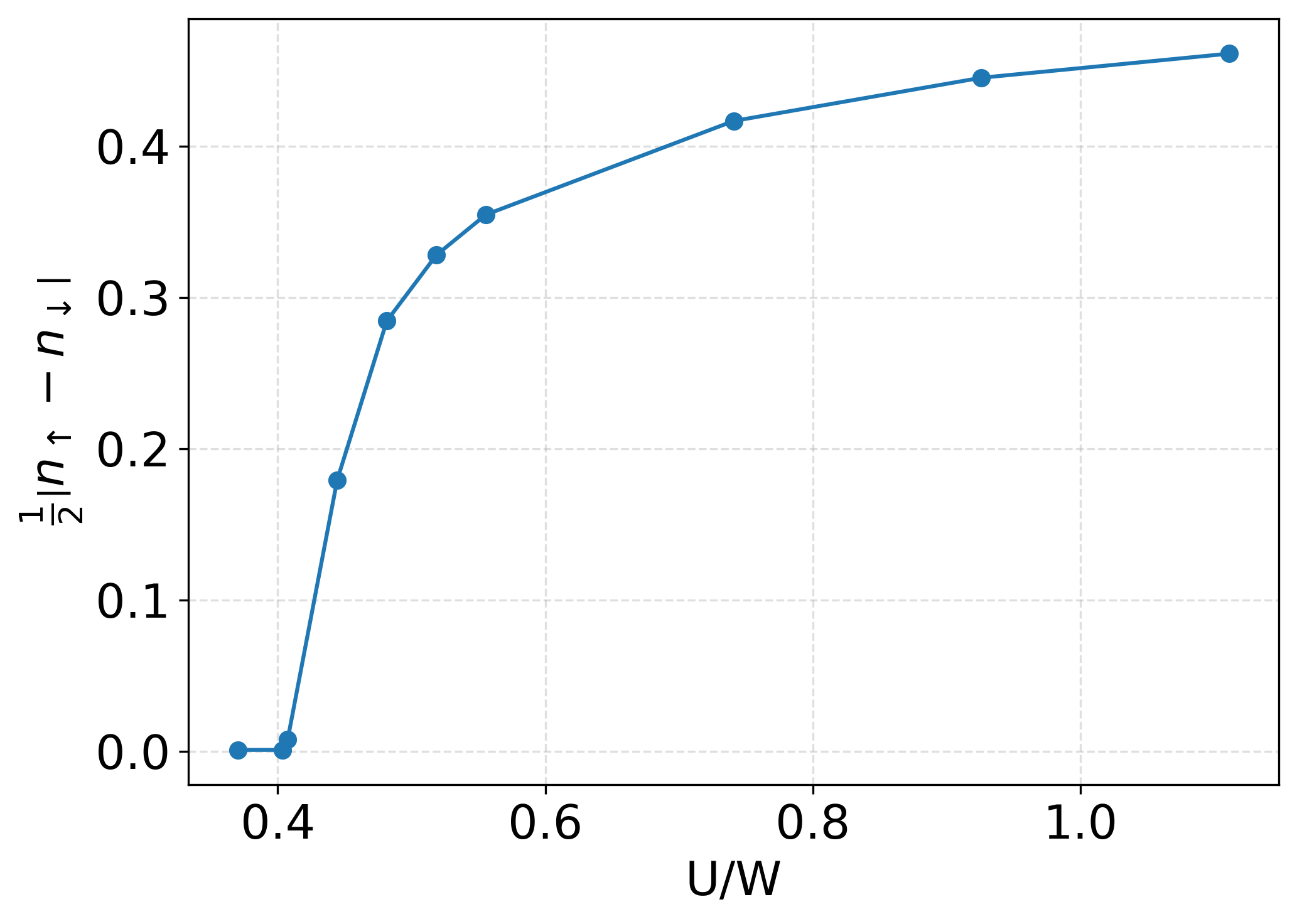}
\caption{
Local magnetization $m=(n_{A\uparrow}-n_{A\downarrow})/2$ as a function of the
interaction strength $U$ at fixed temperature $T/W=0.001$. The magnetization
decreases continuously with decreasing $U$ and vanishes at a critical value
$U_c/W \approx 0.4$, indicating the transition out of the magnetically ordered
phase.
\label{fig_mag_U}
}
\end{figure}
We now turn to the dependence of the magnetic state on the interaction strength
$U$. Figure~\ref{fig_mag_U} shows the local magnetization on sublattice $A$,
$m\equiv m_A=(n_{A\uparrow}-n_{A\downarrow})/2$, as a function of $U$ at the
fixed temperature $T/W=0.001$. The two sublattices carry exactly opposite
magnetizations, $m_B=-m_A$, so that the total magnetization vanishes. As $U$ is
reduced, $m$ decreases continuously and eventually vanishes at
$U_c/W\approx0.4$, indicating the disappearance of the magnetically ordered
phase.

\begin{figure*}[t]
\centering 
\begin{overpic}[width=0.32\linewidth]{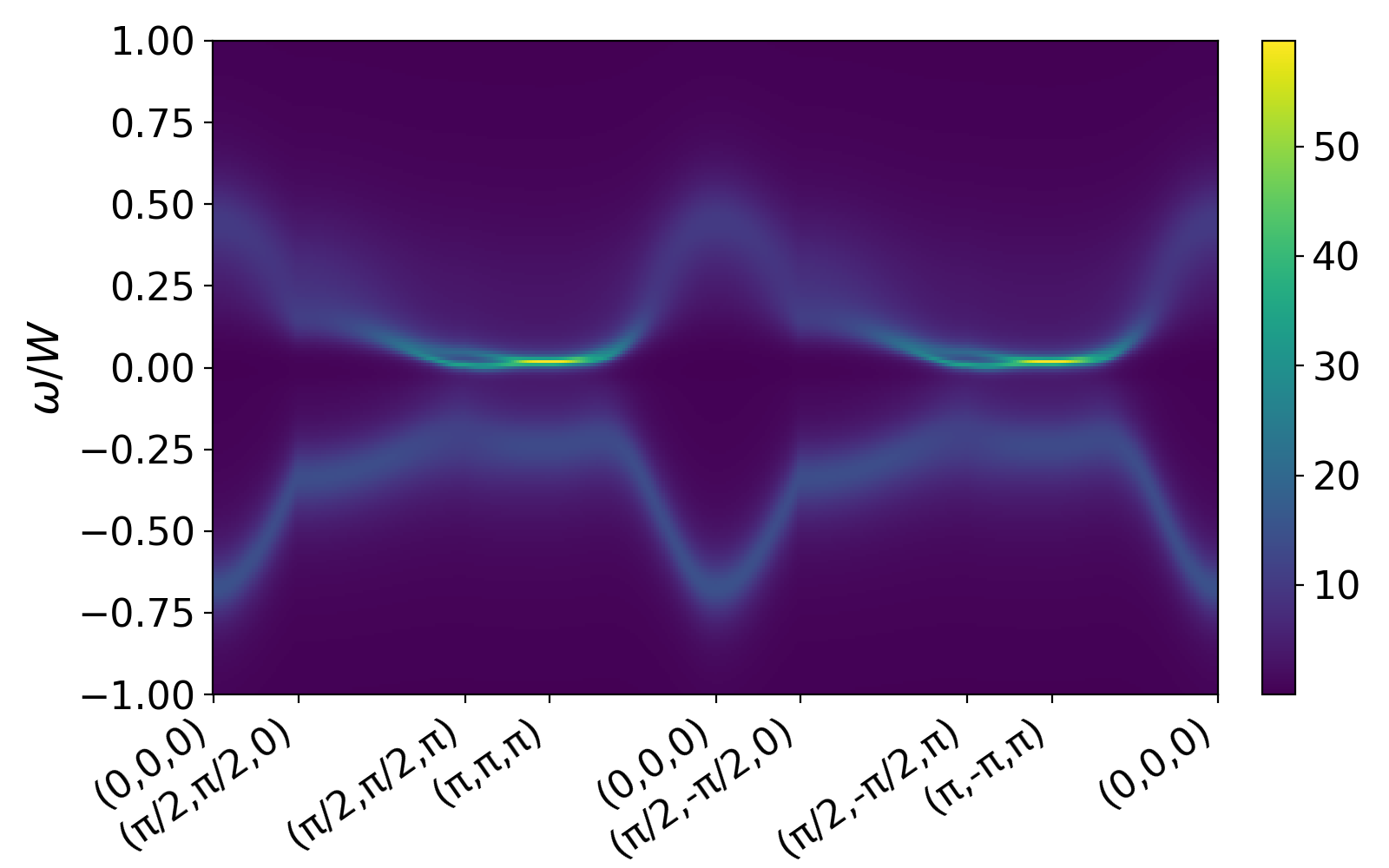}
\put(-4,60){(a)} 
\end{overpic} 
\hfill 
\begin{overpic}[width=0.32\linewidth]{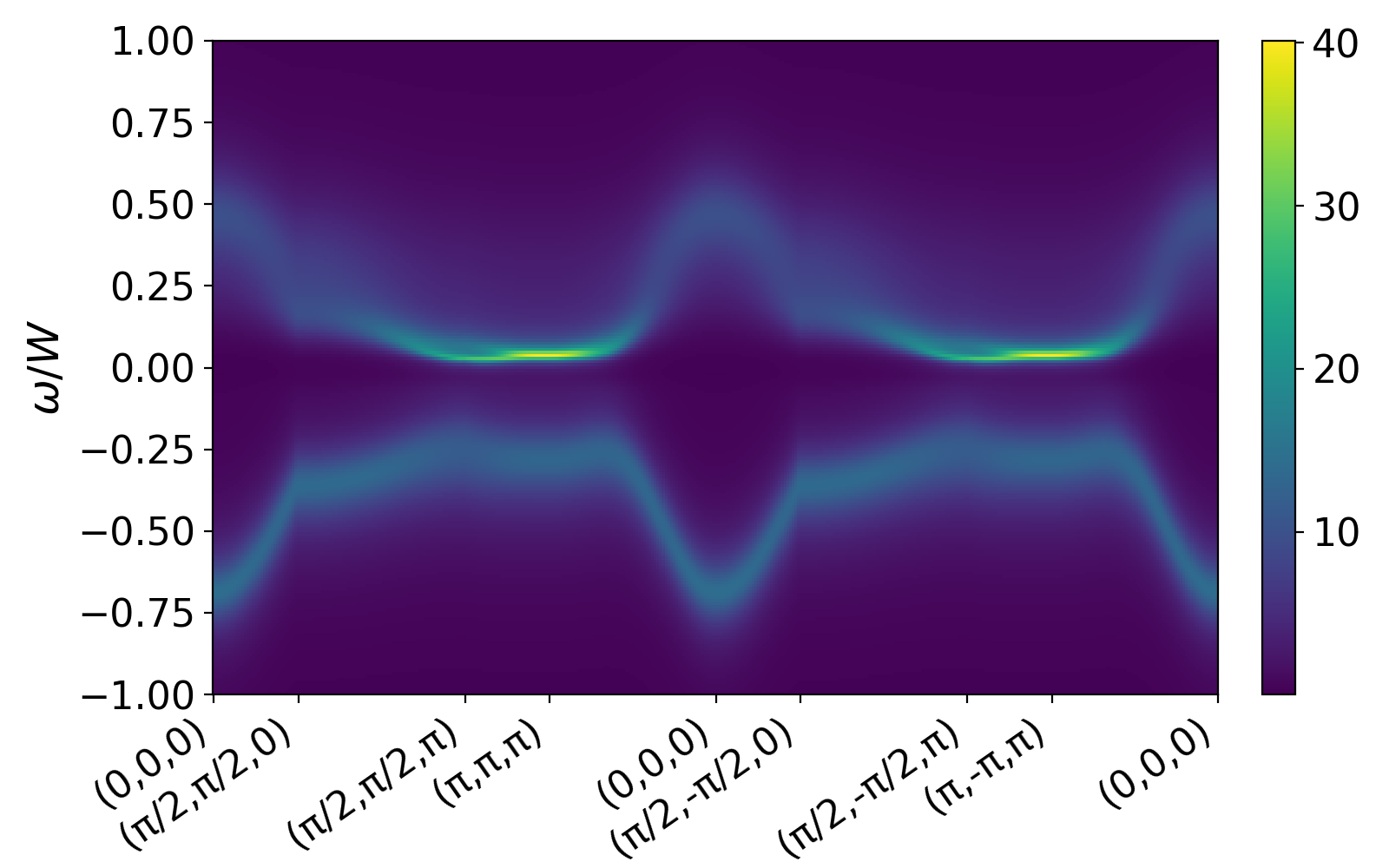}
\put(-4,60){(b)} 
\end{overpic} 
\hfill 
\begin{overpic}[width=0.32\linewidth]{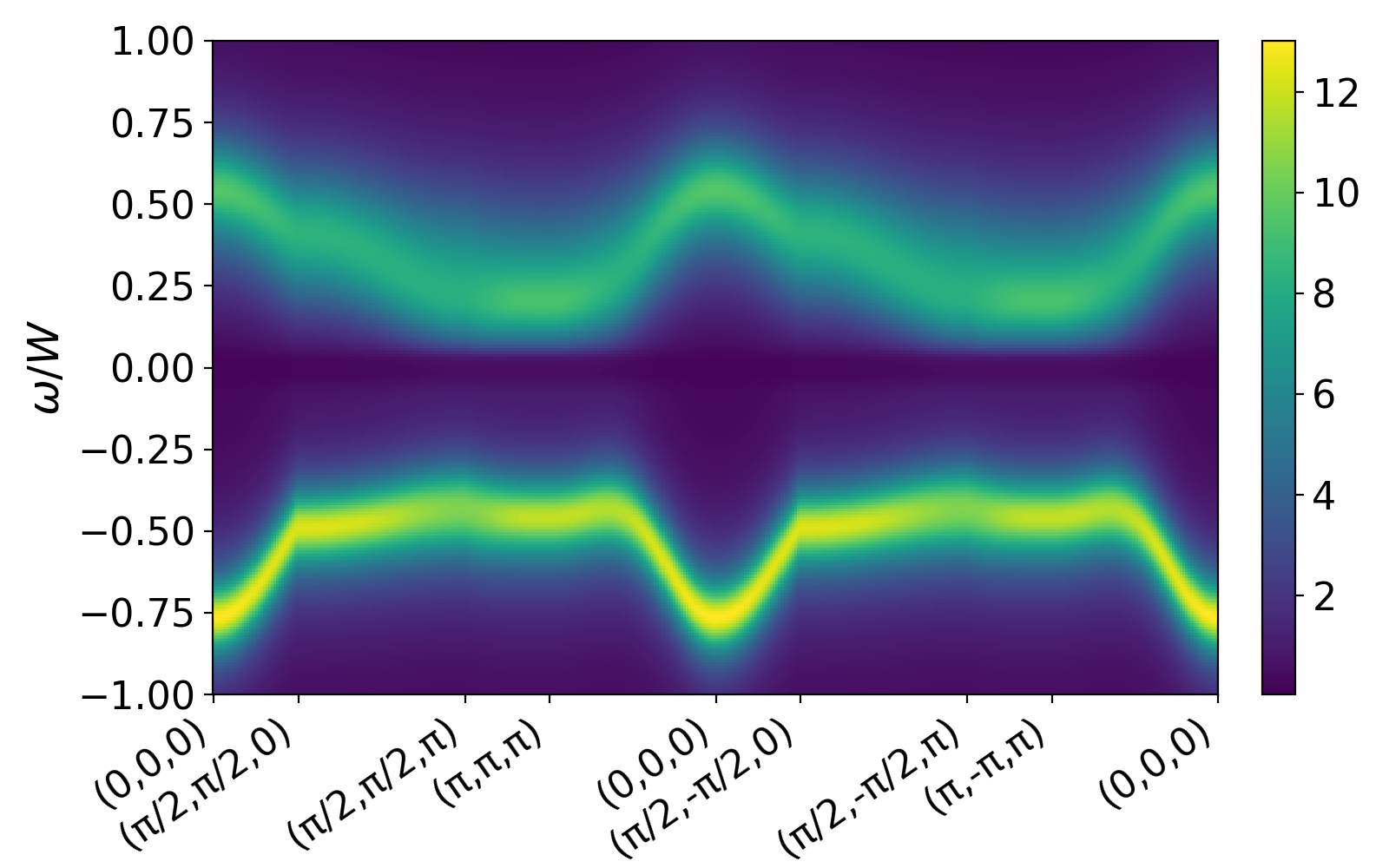}
\put(-4,60){(c)} 
\end{overpic} 
\hfill 
\begin{overpic}[width=0.32\linewidth]{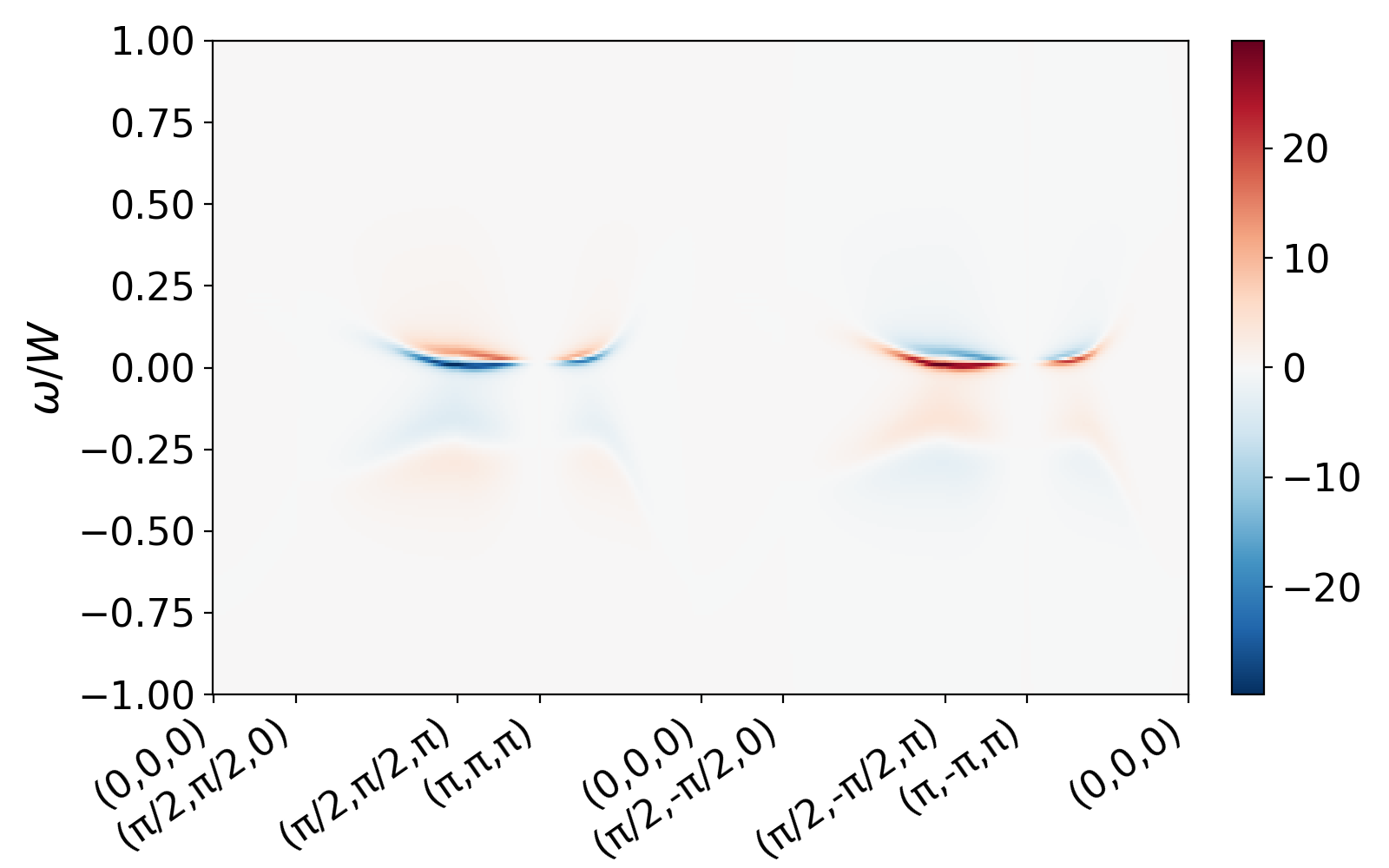}
\put(-4,60){(d)} 
\end{overpic} 
\hfill 
\begin{overpic}[width=0.32\linewidth]{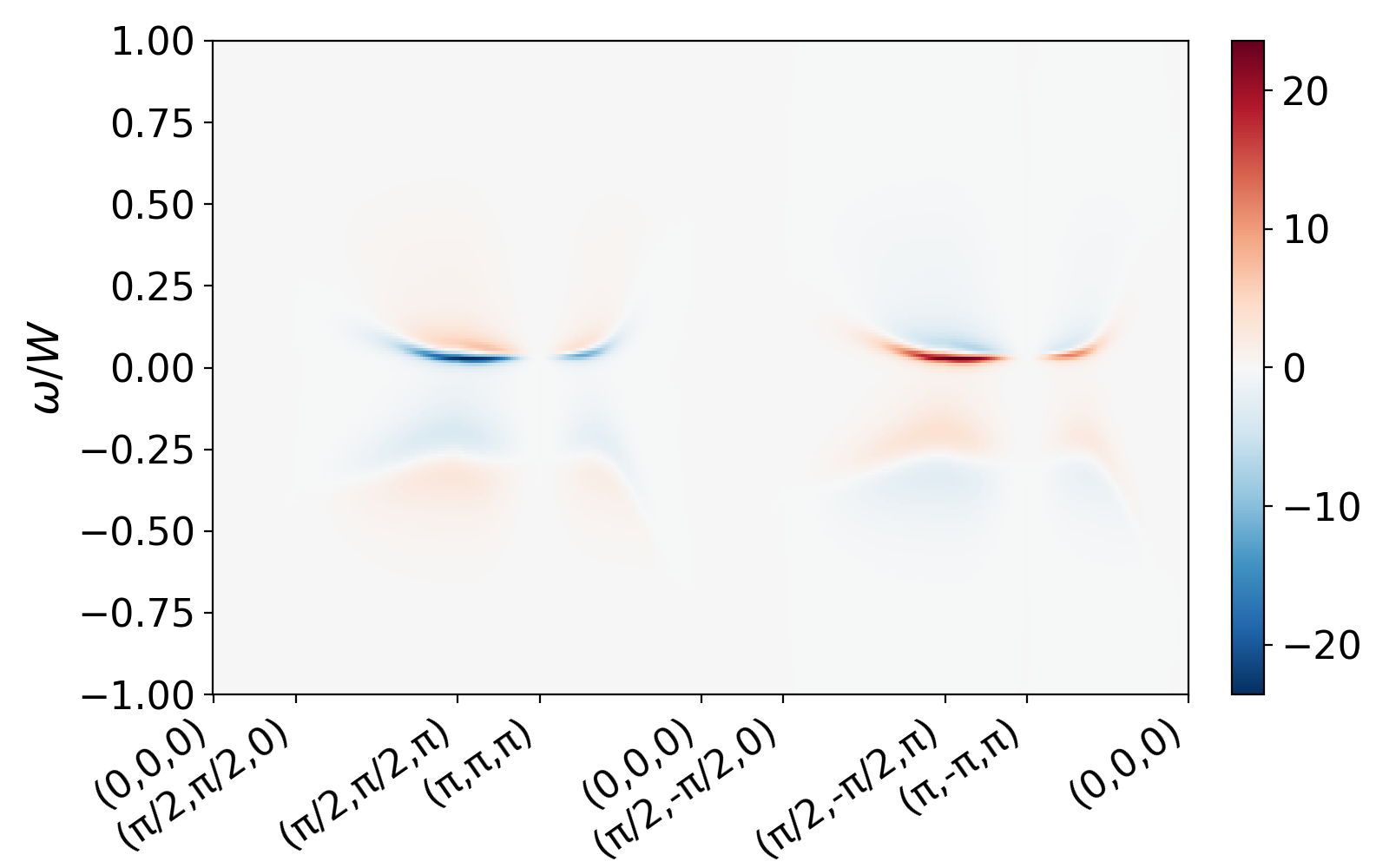}
\put(-4,60){(e)} 
\end{overpic} 
\hfill 
\begin{overpic}[width=0.32\linewidth]{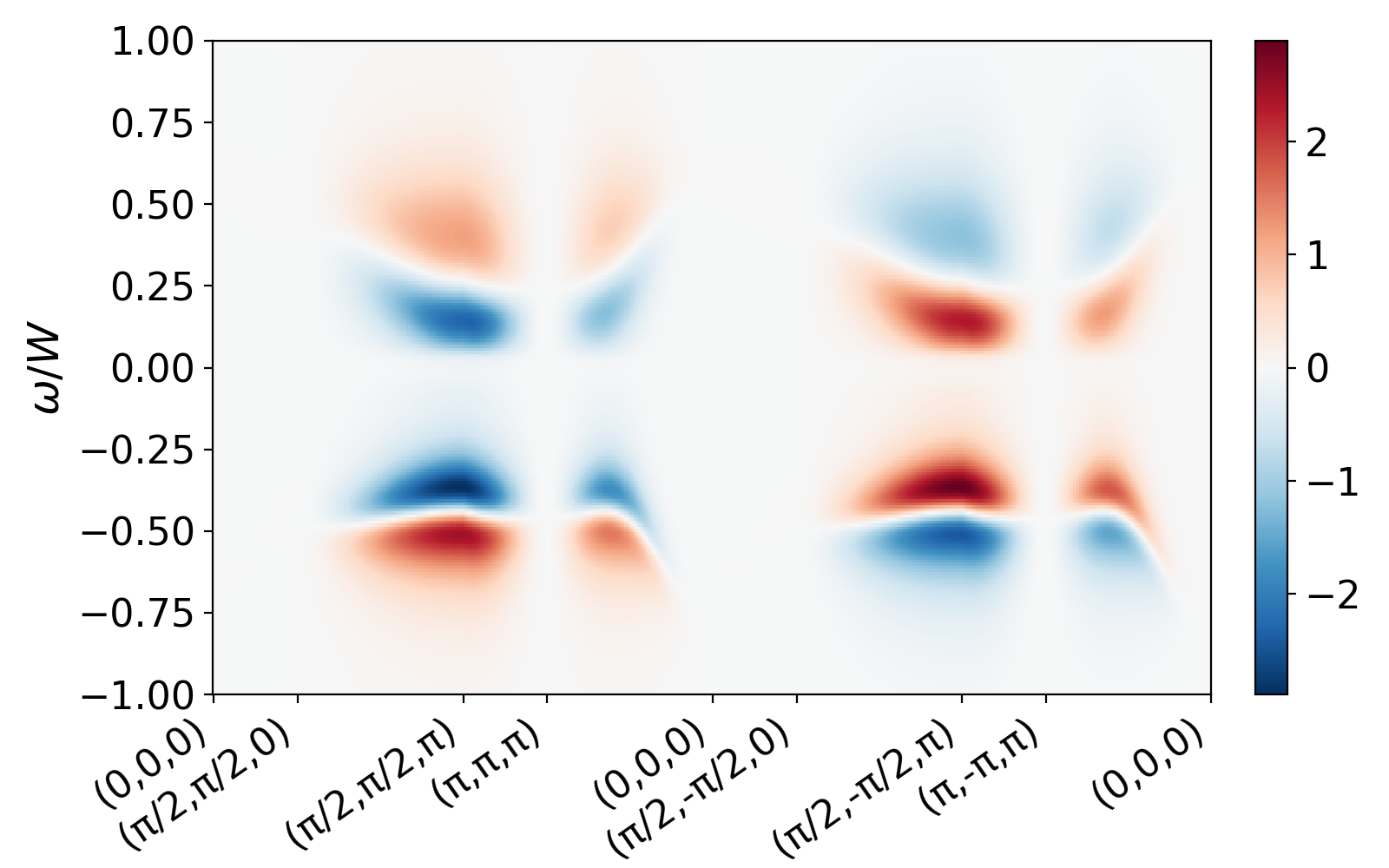}
\put(-4,60){(f)} 
\end{overpic} 
\caption{
Spectral evolution with interaction strength $U/W$.
(a)--(c): Total spectral function
$A_\uparrow + A_\downarrow$ for $U/W=0.55$, $0.63$, and $0.92$,
from left to right.
(d)--(f): Spin-resolved spectral difference
$A_\uparrow - A_\downarrow$ for the same values of $U/W$.
While the overall spectral structure remains similar, the gap at the Fermi
energy increases with increasing $U/W$. The spin-resolved spectral difference
exhibits the characteristic $d$-wave-like altermagnetic pattern for all
interaction strengths shown. We note that the color scale is chosen separately
for each panel; therefore, the apparent color intensity should not be compared
directly between different values of $U/W$. In absolute magnitude, as indicated
by the corresponding color bars, the spin-resolved spectral contrast is largest
for $U/W=0.55$ and is reduced as $U/W$ increases.
\label{fig_Spec_diffU}
}
\end{figure*}

The evolution of the spectral properties with $U$ is shown in
Fig.~\ref{fig_Spec_diffU}. The total spectral function
$A_\uparrow + A_\downarrow$ retains the same overall structure as in the
reference case, while the most prominent change is the increase of the gap at
the Fermi energy with increasing $U$. The lower panels show the spin-resolved
spectral difference $A_\uparrow - A_\downarrow$ along representative momentum
cuts. For all values of $U$ inside the ordered phase, the spectra exhibit the
characteristic sign-changing pattern in the $k_x$--$k_y$ plane, consistent with
the underlying altermagnetic order.

Interestingly, the spin-resolved spectral contrast appears more pronounced at
smaller $U$, even though the local magnetization increases with increasing
interaction strength. This apparent contrast reflects the fact that
$A_\uparrow - A_\downarrow$ is a momentum- and energy-resolved quantity rather
than a direct measure of the local magnetic moment. In the present
altermagnetic state, the momentum-resolved spectral function contains
contributions from both sublattices, giving rise to the sign-changing
$d$-wave-like structure of the spin polarization in momentum space.
The reduction of the spin contrast at larger $U$ can be understood from the
DMFT spectral function. Increasing $U$ shifts the spin-polarized features to
higher energies and enhances the imaginary part of the self-energy away from
the Fermi level. The resulting broadening causes the spin-resolved spectral
features to overlap more strongly in energy, thereby reducing the magnitude of
$A_\uparrow(\omega)-A_\downarrow(\omega)$ at a fixed energy despite the larger
local magnetization.

\begin{figure}[t]
\includegraphics[width=\linewidth]{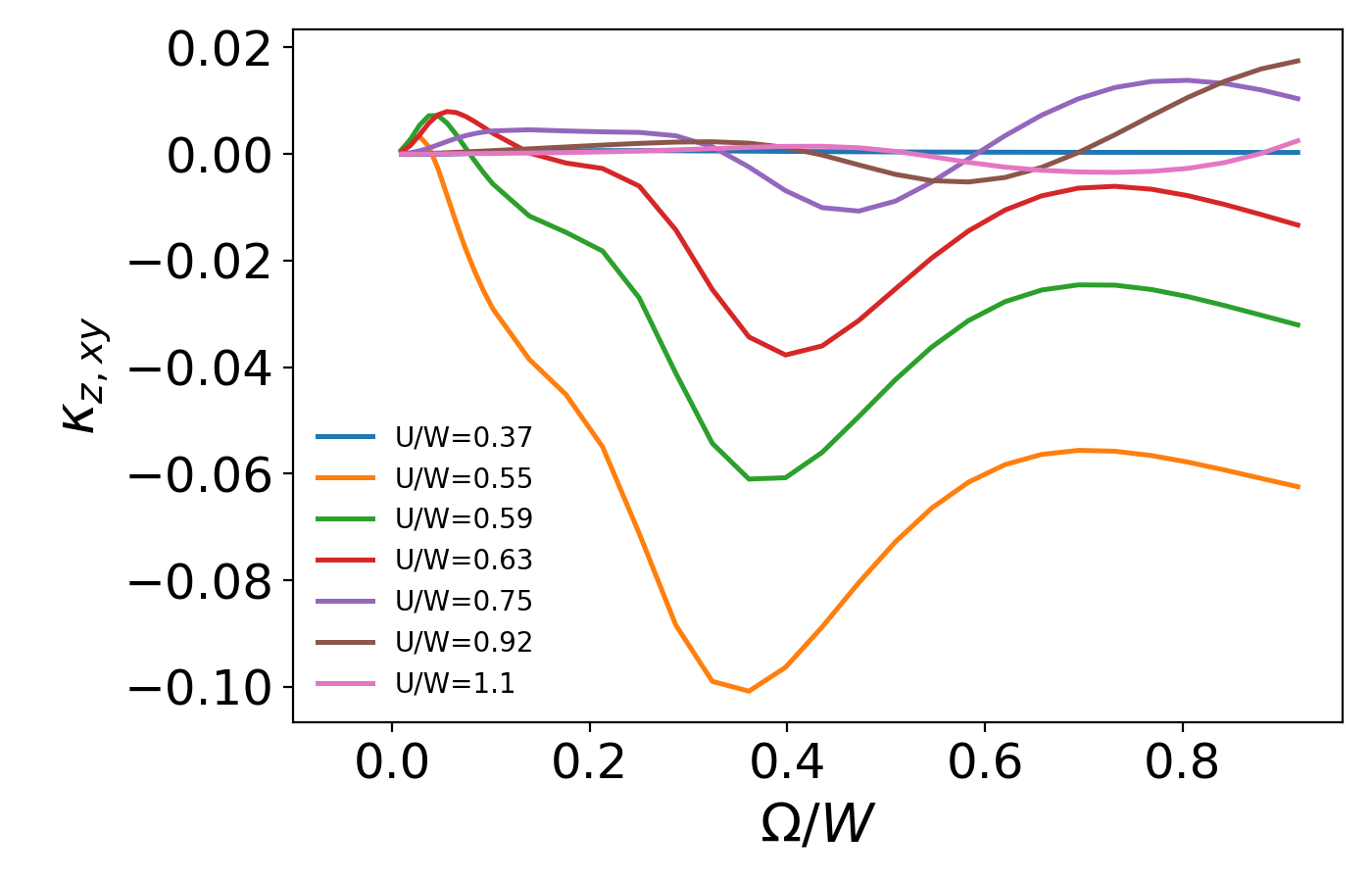}
\caption{
NMEE $\kappa_{z,xy}(\Omega)$ for different interaction
strengths $U/W$, calculated at fixed broadening $\eta/W=0.036$.
For $U/W>U_c/W$, the overall structure of $\kappa_{z,xy}(\Omega)$ remains
qualitatively similar, while the peak positions shift as the characteristic
spectral energy scales evolve with interaction strength. In contrast, for
$U/W<U_c/W$ the response vanishes in the paramagnetic phase. The magnitude of
the response is enhanced as $U/W$ approaches the phase boundary from the ordered
side.
\label{fig_NMEE_U}}
\end{figure}
\begin{figure}[t]
    \centering
    \includegraphics[width=\columnwidth]{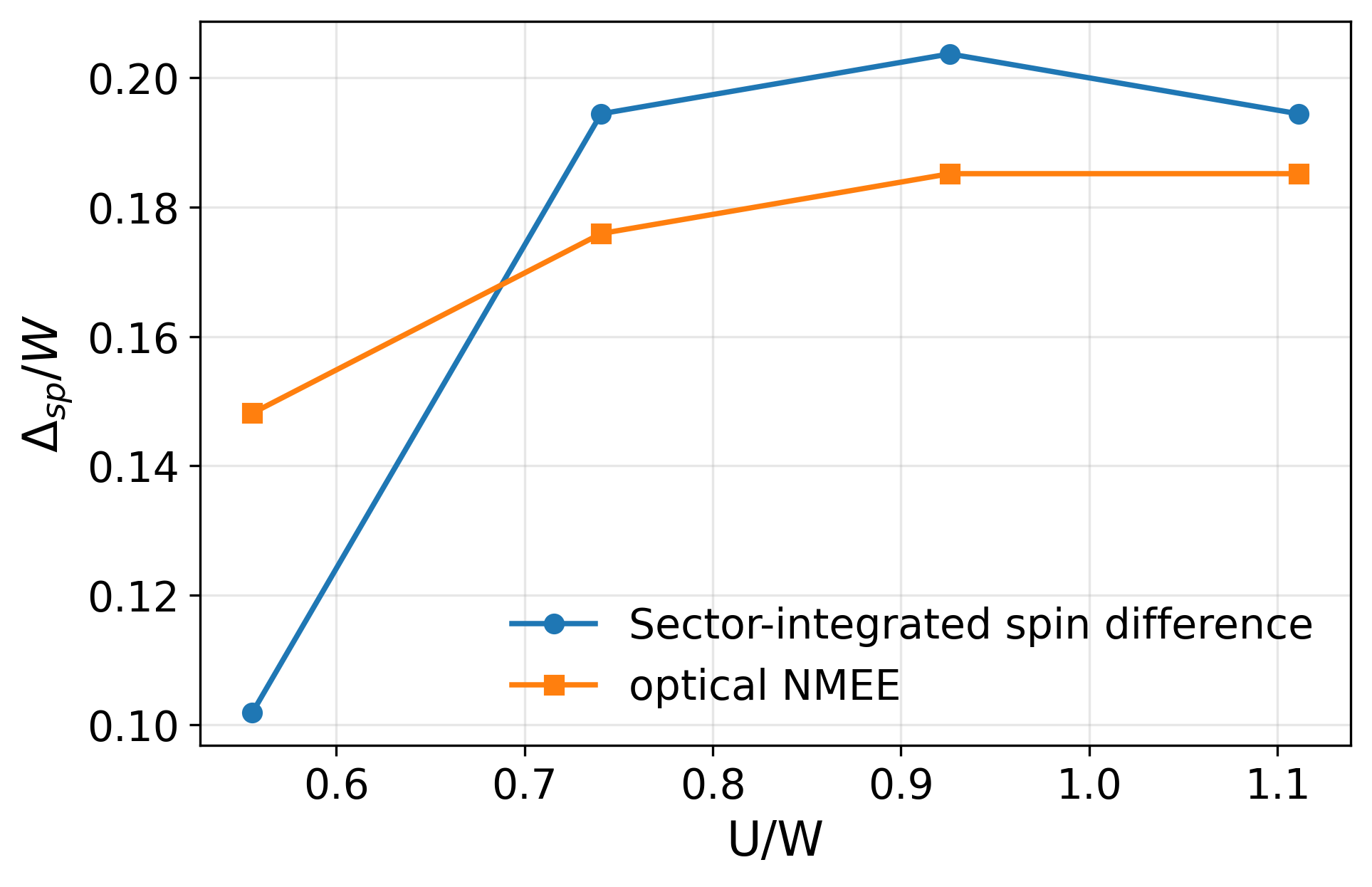}
    \caption{
    Interaction dependence of the characteristic altermagnetic spin-splitting
    scale extracted from the single-particle spectrum and from the optical
    NMEE. The spectral estimate,
    $\overline{\Delta}_{\mathrm{sp}}
    =(\Delta_{\mathrm{sp}}^{\mathrm{occ}}
    +\Delta_{\mathrm{sp}}^{\mathrm{unocc}})/2$,
    is obtained from the peak separations of the spin-difference spectrum
    integrated over one quarter of the Brillouin zone. The optical estimate is
    obtained as $\overline{\Delta}_{\mathrm{sp}}^{\mathrm{NMEE}}
    =\Delta\Omega/2$, where $\Delta\Omega$ is the separation between the two
    higher-frequency extrema of
    $\operatorname{Re}\kappa_{z,xy}(\Omega)$.    (Lines are guides to the eye.) }
    \label{fig_spin_splitting_comparison}
\end{figure}

With the spectral properties established above, we now analyze the interaction
dependence of the NMEE, shown in
Fig.~\ref{fig_NMEE_U}. For interaction strengths within the altermagnetic phase
($U/W>U_c/W$), the overall structure of $\kappa_{z,xy}(\Omega)$ remains qualitatively
similar. In particular, the characteristic peak structure persists, reflecting
the underlying spin-polarized spectral features. As $U/W$ is varied, the peak
positions shift in frequency, consistent with the evolution of the relevant
spectral energy scales, most notably the increase of the gap at the Fermi
energy with increasing $U/W$. In contrast, once the system enters the
paramagnetic phase ($U/W<U_c/W$), the nonlinear response vanishes, consistent
with the absence of spin polarization required for a finite NMEE.

A key observation is that the magnitude of $\kappa_{z,xy}(\Omega)$ is enhanced
throughout the frequency range as $U/W$ approaches the phase boundary from the
ordered side. This behavior follows the evolution of the spin-resolved spectral
contrast in the altermagnetic phase. As discussed above, smaller $U/W$ leads to
a more pronounced momentum- and energy-resolved spin contrast, even though the
local magnetization is reduced. The largest spin contrast occurs close to the
Fermi energy, where the gap is also reduced. Consequently, transitions
involving low-energy states, in particular from occupied states near the Fermi
level to spin-polarized spectral weight slightly above it, make an especially
strong contribution to the nonlinear response.

With increasing $U/W$, by contrast, the spin-polarized spectral weight is
shifted away from the Fermi level and distributed over a broader energy range.
At the same time, the imaginary part of the DMFT self-energy increases away
from the Fermi level, leading to stronger lifetime broadening of the spectral
features. These effects reduce the effective spin-resolved spectral contrast
entering the response kernel and suppress the peak amplitudes of
$\kappa_{z,xy}(\Omega)$. Thus, the response is not controlled only by the
characteristic transition energies, but also by the amount, sharpness, and
energy distribution of the spin-polarized spectral weight participating in the
corresponding transitions.

Finally, Fig.~\ref{fig_spin_splitting_comparison} compares the characteristic
spin-splitting scale extracted from the sector-integrated spectrum with the
estimate obtained from half the separation of the two higher-frequency NMEE
extrema. Over the investigated interaction range, both procedures yield
comparable energy scales, supporting the interpretation of the NMEE peak
separation as a semiquantitative optical estimate of the altermagnetic spin
splitting. The largest deviation occurs at the smallest interaction strength.
This deviation reflects the different sensitivity of the two estimates to the
low-energy spectral renormalization. As seen in
Fig.~\ref{fig_Spec_diffU}(d), the unoccupied spin-polarized features then lie
close to the Fermi energy and are strongly renormalized. The peak-based
spectral estimate is therefore reduced substantially through the small
unoccupied peak separation, even though the occupied spin-polarized features
retain a similar separation. The NMEE estimate, by contrast, is extracted from
finite-frequency extrema that represent momentum- and matrix-element-weighted
transition manifolds rather than direct differences between individual spectral
maxima. It is therefore less tied to the precise positions of the strongly
renormalized unoccupied peaks. This explains why the two estimates remain on
the same characteristic scale but show their largest quantitative discrepancy
at small $U/W$.

\subsection{Temperature dependence of the magnetic phase and nonlinear response}
\label{subsec:T_dependence}
\begin{figure}[t]
\includegraphics[width=\linewidth]{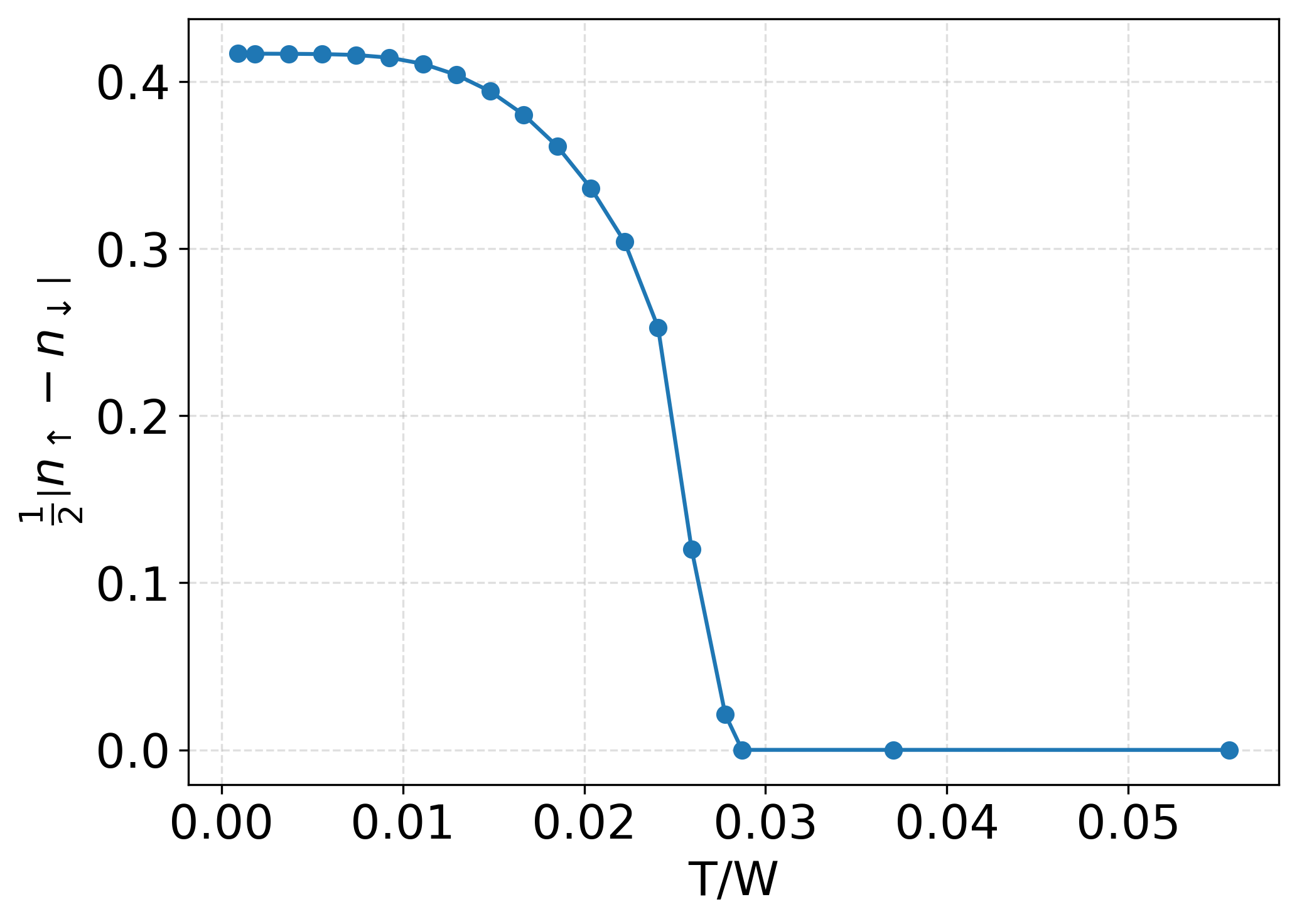}
\caption{
Local magnetization $m=(n_{A\uparrow}-n_{A\downarrow})/2$ as a function of
temperature $T/W$ at fixed interaction strength $U/W=0.75$. The magnetization decreases
with increasing temperature and vanishes at a critical temperature
$T_c/W \approx 0.028$, indicating the transition from the 
altermagnetic phase to the paramagnetic phase.
\label{fig_mag_T}
}
\end{figure}

\begin{figure*}[t]
\includegraphics[width=0.31\linewidth]{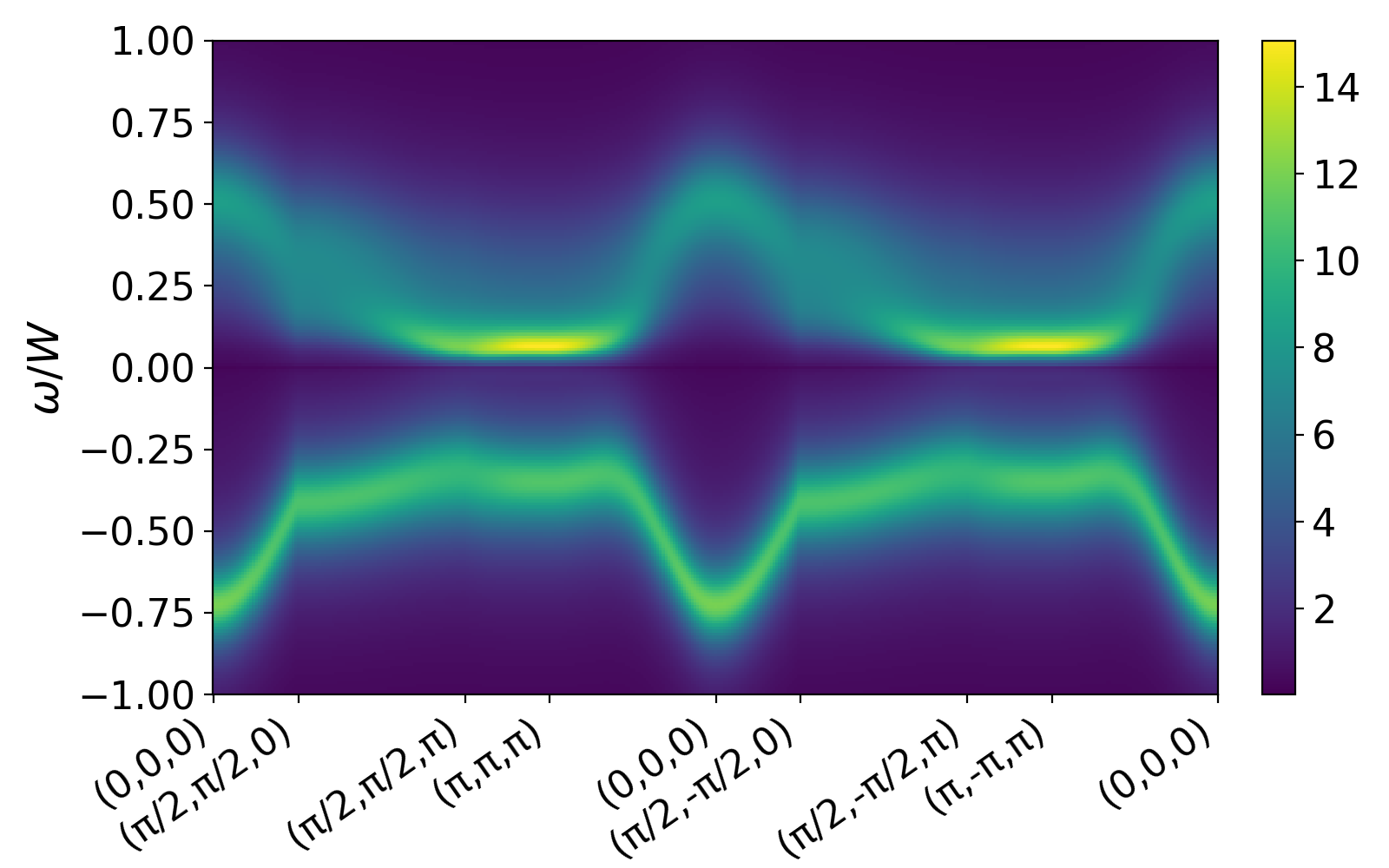}
\includegraphics[width=0.31\linewidth]{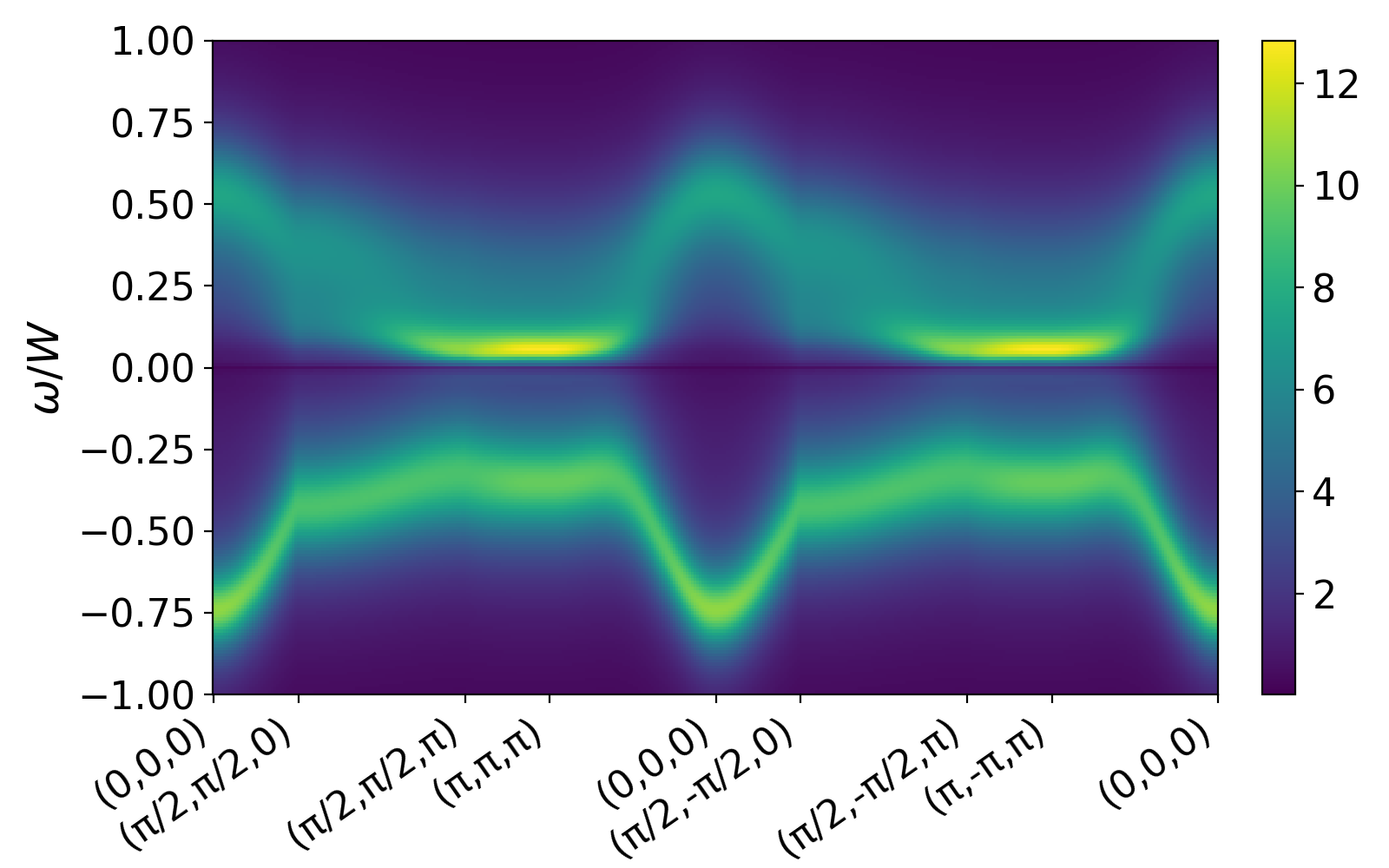}
\includegraphics[width=0.31\linewidth]{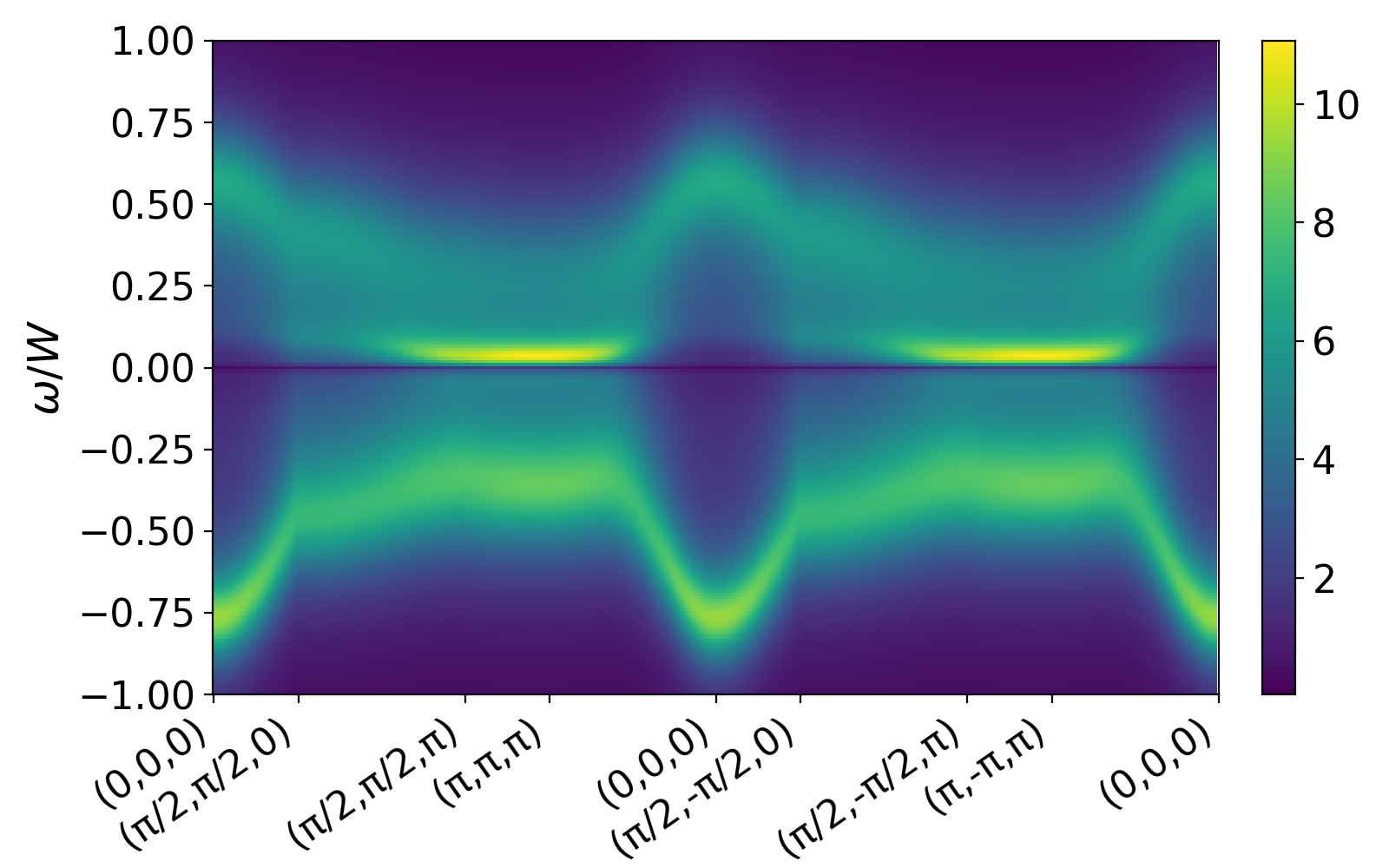}
\includegraphics[width=0.31\linewidth]{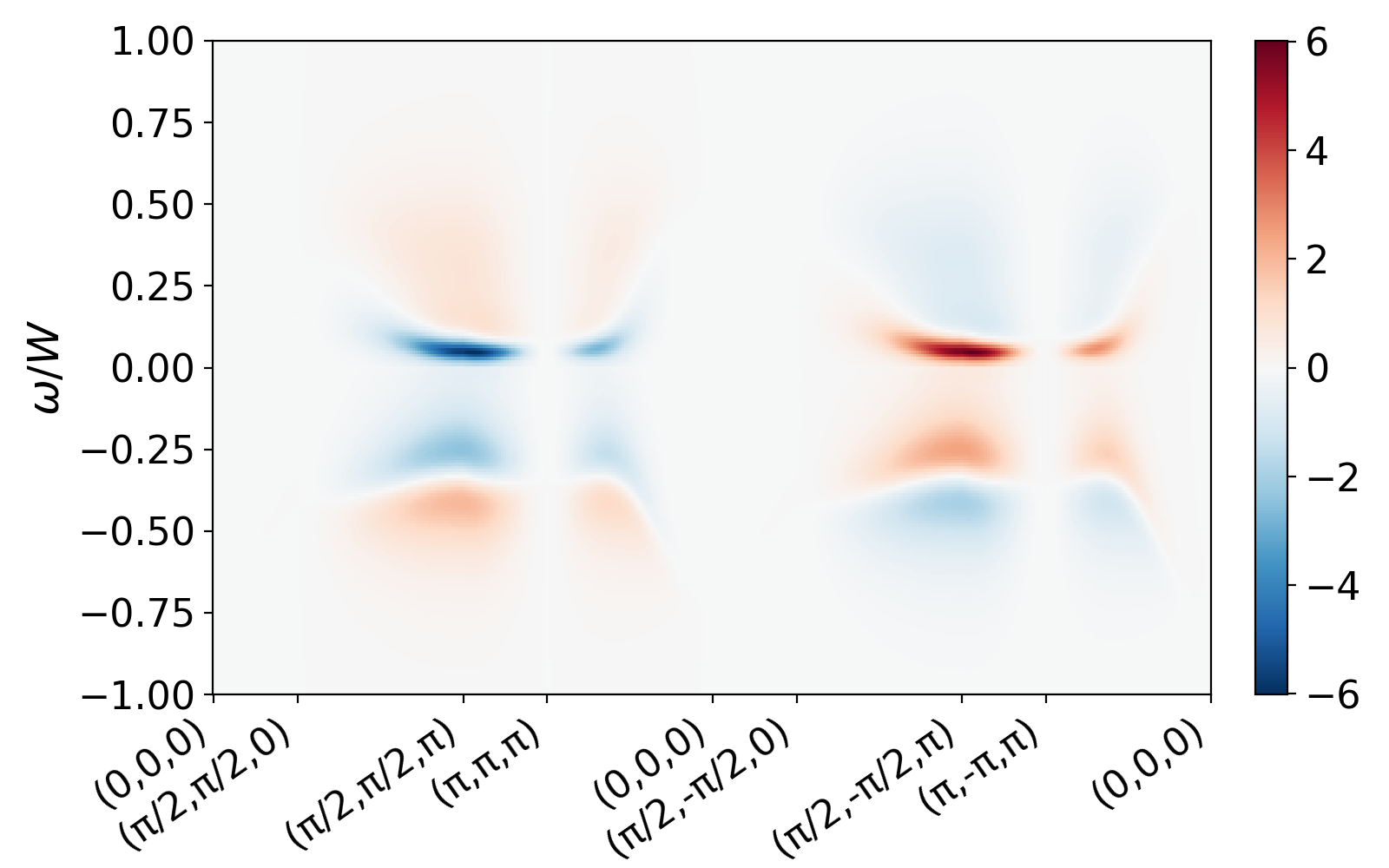}
\includegraphics[width=0.31\linewidth]{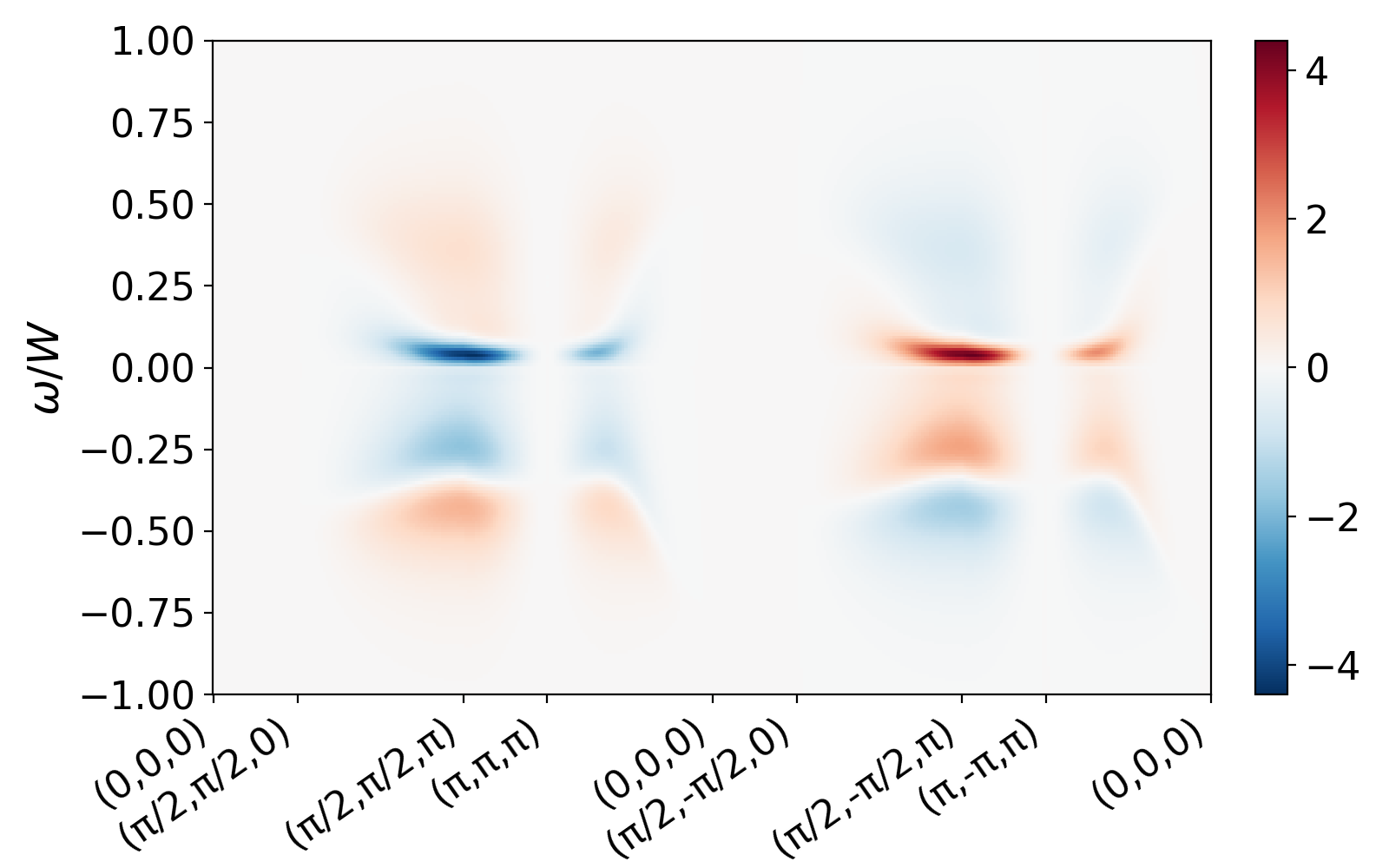}
\includegraphics[width=0.31\linewidth]{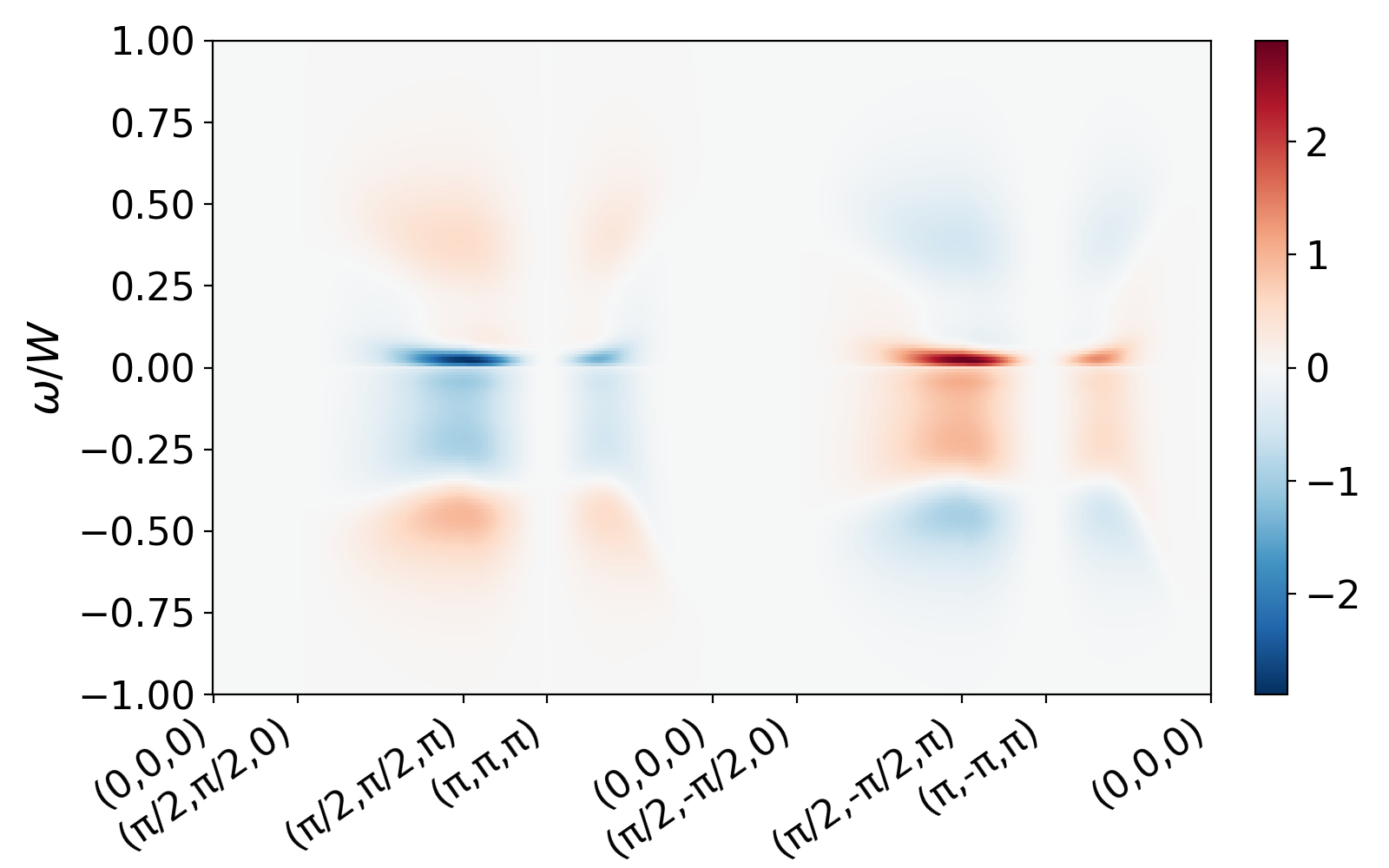}
\caption{
Temperature evolution of the spectral functions and spin-resolved spectral
contrast at fixed interaction strength $U/W=0.75$.
From left to right, the panels correspond to $T/W=0.016$, $0.020$, and $0.024$.
With increasing temperature, the spectral features become broader, the gap at
the Fermi energy is gradually reduced, and the amplitude of the spin-resolved
spectral difference decreases. This reflects the weakening of the
altermagnetic order as the system approaches the temperature-driven transition.
\label{fig_Spec_diffT}
}
\end{figure*}
\begin{figure}[t]
\includegraphics[width=\linewidth]{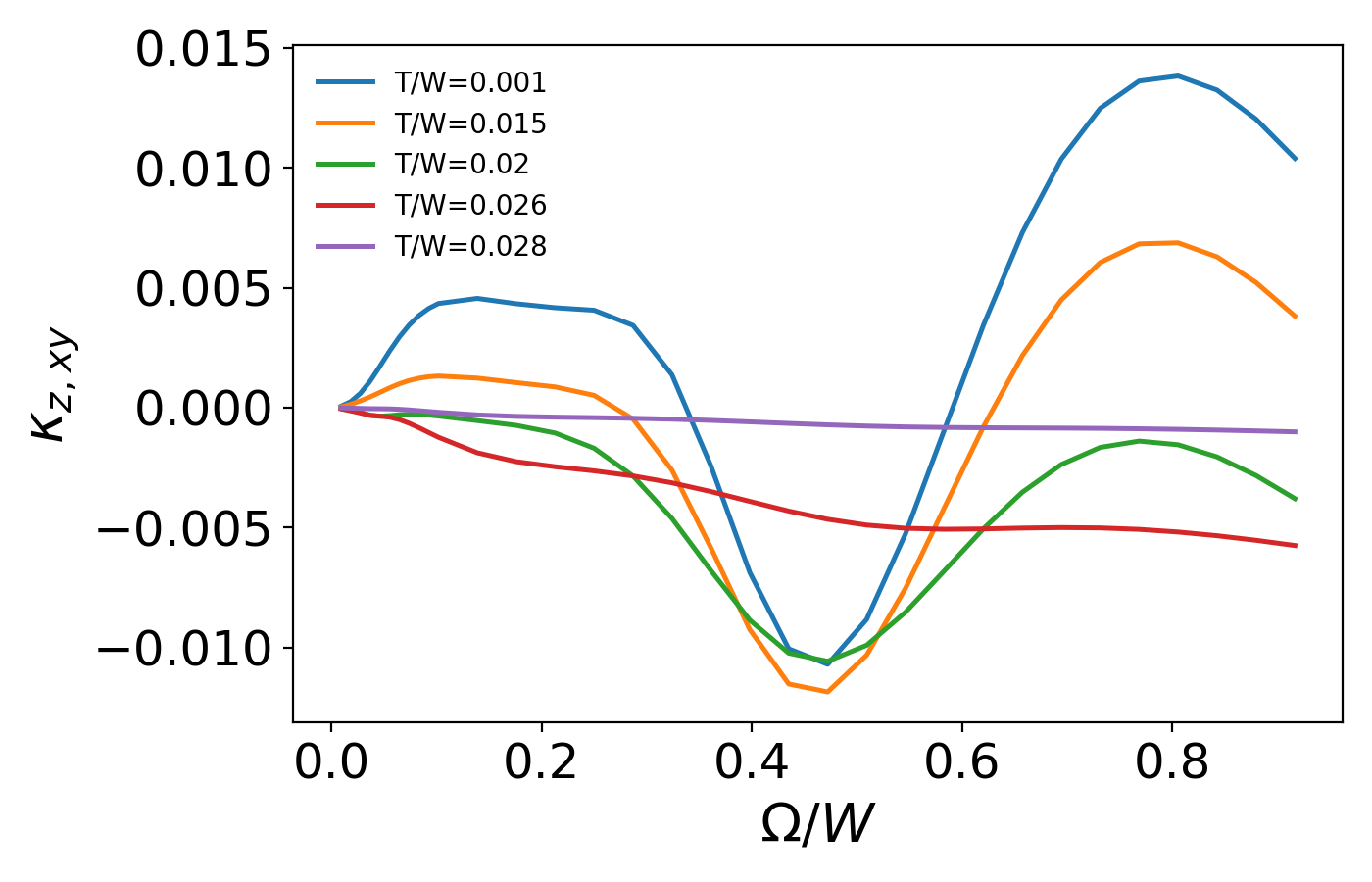}
\caption{
Temperature dependence of the NMEE
$\kappa_{z,xy}(\Omega)$ at fixed interaction strength $U/W=0.75$. In contrast to the interaction-driven
evolution, increasing temperature does not enhance the response near the phase
boundary. Instead, the overall response is suppressed as the system approaches
the critical temperature $T_c/W \approx 0.028$ and vanishes in the
paramagnetic phase. 
\label{fig_NMEE_T}
}
\end{figure}
We now turn to the temperature-driven phase transition at fixed interaction
strength $U/W=0.75$. As the temperature is increased, the system undergoes a
transition from the altermagnetic phase to a paramagnetic state.

A clear signature of this transition is the suppression of the local
magnetization $m=(n_{A\uparrow}-n_{A\downarrow})/2$. As shown in
Fig.~\ref{fig_mag_T}, $m$ decreases continuously with increasing temperature
and vanishes at $T_c/W \approx 0.028$, indicating the loss of magnetic order.

The evolution of the spectral function, shown in Fig.~\ref{fig_Spec_diffT}, provides further insight into the nature of this transition. With increasing temperature, the imaginary part of the self-energy grows, leading to a progressive broadening of the spectral features. As a consequence, spectral weight is redistributed from energies above and below the Fermi level toward the Fermi energy, resulting in a gradual closing of the gap. At the same time, the spin-resolved spectral functions become increasingly
similar, and the spin-resolved spectral contrast
$A_\uparrow(\omega)-A_\downarrow(\omega)$ is reduced as the transition is
approached.
This behavior differs qualitatively from the interaction-driven transition discussed above. While increasing $U$ enhances the local magnetization but reduces the spectral spin contrast through a redistribution of spectral weight, increasing temperature suppresses both the magnetic order and the spin-resolved spectral contrast simultaneously. In this case, the reduction of the spin-resolved spectral contrast is directly tied to the loss of magnetic order and the increasing scattering encoded in the self-energy.

We finally analyze the temperature dependence of the NMEE at fixed interaction strength $U/W=0.75$, shown in
Fig.~\ref{fig_NMEE_T}.  In contrast to the interaction-driven evolution
discussed above, increasing temperature does not lead to an enhancement of the
response close to the phase boundary. Instead, the overall structure of
$\kappa_{z,xy}(\Omega)$ flattens and is rather suppressed as the temperature is
increased and the altermagnetic order weakens. Above the critical temperature
$T_c/W \approx 0.028$, the response vanishes, consistent with the disappearance
of the spin polarization required for a finite NMEE.

The suppression is, however, not a simple uniform reduction of the entire
spectrum. The negative peak around $\Omega/W \simeq 0.4$ remains comparatively
robust throughout the ordered phase and changes only moderately with
temperature. By contrast, the positive structures at low frequencies and at
higher frequencies around $\Omega/W \simeq 0.75$--$0.8$ are strongly reduced
with increasing temperature. At temperatures close to the transition, these
contributions even change sign and become negative. This indicates that the
temperature dependence of $\kappa_{z,xy}(\Omega)$ is governed not only by the
reduction of the magnetic order parameter, but also by a redistribution of the
relative weights of different transition channels contributing with opposite
signs.

This behavior is consistent with the temperature evolution of the spectral
functions discussed above. As temperature increases, the spectral features
broaden, the gap near the Fermi level is reduced, and the spin-resolved
spectral contrast decreases. These effects suppress the spin-polarized
spectral weight available for the nonlinear response. At the same time, thermal
broadening modifies the balance between positive and negative contributions
from different momentum and energy regions. The resulting response therefore
does not scale simply with the magnetization, but reflects the detailed
temperature-dependent redistribution of spin-polarized spectral weight in the
altermagnetic phase.

\section{Summary and Conclusions}
\label{sec:conclusion}
In this work, we have investigated the NMEE
in a correlated altermagnetic system using DMFT. In
contrast to effective band descriptions with an imposed spin splitting, the
magnetic order, spectral function, and nonlinear response are determined
self-consistently. This allows us to directly relate the optical NMEE to the
spin-resolved electronic structure of the correlated altermagnetic phase.

We find that the NMEE is finite in the altermagnetic phase and vanishes in the
paramagnetic phase. Its frequency dependence reflects the spin-polarized
spectral features of the altermagnetic state.
 The higher-frequency peak structure of $\kappa_{z,xy}(\Omega)$ further provides
an indirect estimate of the characteristic spin-splitting scale: the separation
between the relevant peaks in the optical NMEE is consistent with approximately twice
the sector-averaged spin splitting extracted from the spin-resolved spectral
function. 
Thus, the optical NMEE not only serves as a probe of altermagnetic order, but
also carries information about the energy scale of the spin-split electronic
structure. In this respect, it provides a complementary bulk-sensitive probe
to spin-resolved ARPES, whose surface sensitivity can complicate the
identification of bulk altermagnetic spin splitting. This quantitative connection between a
bulk nonlinear response and the spin-splitting scale is a distinctive feature
of the present analysis. The magnitude of the response is controlled not
simply by the local magnetization, but by the momentum- and energy-resolved
spin contrast in the spectral function.

The interaction and temperature dependences show qualitatively different
behavior. When $U/W$ is reduced toward the magnetic phase boundary at low
temperature, the NMEE is enhanced. This behavior is particularly
interesting because this interaction-driven transition can be viewed as the
low-temperature continuation of an underlying quantum phase transition
between the altermagnetic and paramagnetic phases. The enhancement originates
from the reduction of the gap and the increased spin-polarized spectral weight
near the Fermi level, which amplify the low-energy contribution to the
response. From an experimental perspective, this suggests that tuning an
altermagnet close to such a phase boundary, for example by pressure, strain, or
chemical substitution, may provide a favorable regime for observing the NMEE.
By contrast, increasing temperature suppresses the response: thermal broadening
and the weakening of altermagnetic order reduce the spin-resolved spectral
contrast, and the NMEE vanishes above $T_c$.

These results demonstrate that the NMEE is a sensitive probe of correlated
altermagnetic order and its underlying spin-resolved spectral properties. From
an experimental standpoint, promising platforms are correlated altermagnetic
insulators or semimetals in which the interaction strength can be tuned toward
a magnetic phase boundary by pressure, strain, or chemical substitution,
analogous to the pressure-driven metal-insulator transition in
NiS$_2$~\cite{pgp6-zlh8} or strain engineering in RuO$_2$-based
heterostructures~\cite{Park2026RuO2}. A possible experimental route is a pump--probe measurement, in which a
linearly polarized pump generates a rectified NMEE spin polarization and a
subsequent linearly polarized probe detects it through magneto-optical
rotation.

\begin{acknowledgments}
J.O. was supported by JST SPRING (Grant Number JPMJSP2110), JSPS KAKENHI (Grant Number JP26KJ1450), and the Iwadare Scholarship Foundation. R.P. was supported by JSPS KAKENHI (Grant Number JP23K03300). Part of the calculations were performed using the facilities of the Supercomputer Center, Institute for Solid State Physics, The University of Tokyo.
\end{acknowledgments}

\bibliographystyle{apsrev4-2}
\bibliography{references}

\end{document}